# Future Prospects for Hadron Physics at P̄ANDA

Ulrich Wiedner

Institut f. Experimentalphysik 1, Ruhr-University Bochum, D-44780 Bochum, Germany

**Abstract:**

The P̄ANDA experiment at the new FAIR facility will be the major hadron physics experiment at the end of this decade. It has an ambitious far-reaching physics program that spans the most fascinating topics that are emerging in contemporary hadron physics. The universality of the antiproton annihilation process, with either protons or nuclei as targets, allows physicists to address questions like the structure of glueballs and hybrids; to clarify the nature of the X, Y and Z states; to investigate electromagnetic channels in order to measure form factors of the nucleon; and to provide theory with input with respect to non-perturbative aspects of QCD. The possibility to use different nuclear targets opens the window for charm physics with nuclei or for color transparency studies, as well as for an intensive hypernuclear physics program. Previous experimental experience has clearly demonstrated that the key to success lies in high levels of precision complemented with sophisticated analysis methods, only possible with high statistics in the data set. However, since this puts many critical demands on the detector, P̄ANDA's design has incorporated cutting-edge detector technologies that in some cases have surpassed even the requirements for LHC experiments.

**Contents:**

## Introduction

The $\overline{\text{P}}$ANDA experiment belongs to the third generation of hadron physics experiments, hereby building on the experiences and successes of previous generations. The name is an acronym and stands for Antiproton ($\overline{\text{p}}$) ANnihilations in DArmstadt.

The availability of accelerators and detectors like the bubble chambers led to a prosperous time in particle physics in the 1950's and 1960's with the discovery of many new particles. The vast majority of these particles came to be understood, at least qualitatively, due to the pioneering work of Gell-Mann [1]. After the discovery of the charm quark, the quark model—which is nowadays referred to as the naïve or constituent quark model (NQM)—was generally accepted and could explain most of the hadronic states. Hadrons are composed of quarks that interact themselves via gluons, baryons are hadrons consisting of three quarks, and mesons are hadrons consisting of an antiquark-quark pair.

The hadron physics of the 1970's and 1980's concentrated on two areas: firstly on the spectrum of hadrons containing the relatively heavy charm and bottom quarks, and secondly on smaller detector set-ups dedicated to specific physics goals in the light-quark sector. In the heavy-quark sector theoretical expectations were met that the quarks and antiquarks in mesons could form pseudoscalar and vector states, or should appear in a P-wave configuration. Experiments reported no clearly superfluous or ambiguous hadron for states with charm or bottom quarks.

Where did this success of the NQM come from? In the case of charm and bottom quarks, hadrons consist of bound systems of approximately non-relativistic heavy constituents, whose effective coupling is relatively small. Non-perturbative effects or higher-order corrections can therefore be neglected. Hadron physics and the strong interaction among quarks seemed to be understood in the heavy-quark sector. The situation was, and remains so today, quite a bit different for light mesons and for baryons containing solely light quarks. Here, the high density of states and their broad widths often makes the identification and interpretation of observed signals rather ambiguous.

In the years after the establishment of heavy-quark states, the field of hadron physics was subdivided, according to the methods and tools used, into the following branches:

- hadron spectroscopy
- hadron structure
- interaction of hadrons.

While specialized experiments have been dedicated to each of those subfields over the past 20 years, the $\overline{\text{P}}$ANDA experiment has been especially designed together with the accelerator to address open and burning questions from all of the subfields [1]. Antiproton annihilation has proven in the past to be a universal tool for carrying out such investigations. Advances not only in detector technology and data treatment but also in theoretical understanding will lead to new deep insights into the strong interaction. The results achievable by $\overline{\text{P}}$ANDA might have far-reaching implications for other fields of physics, since an understanding of $\overline{\text{P}}$ANDA physics means dealing with non-perturbative phenomena. Non-perturbative phenomena appear often in physics, but aside from hadron physics, they are hard to study systematically.

# Hadron Spectroscopy with $\overline{\text{P}}$ANDA

**Previous achievements**

In the last century, quantum theory and Maxwell's electrodynamics were integrated into a powerful proper theoretical description called quantum electrodynamics (QED). The experimental discoveries of particles and their properties helped to develop the theoretical framework further, and a common understanding of the weak and the electromagnetic interactions emerged. Building on these successes, the theory of the strong interaction, quantum chromodynamics (QCD), was developed on the basis of the exact color symmetry SU(3) for quarks and their force carriers, the gluons. The Standard Model of particle physics was born.

In the Standard Model, all the forces that make up the different interactions show basically the same behavior, with a force proportional to the inverse-square of the distance. The proper sets of theories are called gauge theories. At this point, one can ask where hadron physics fits in and what is its specific role for our understanding of physics.

When trying to understand composite, strongly interacting particles two different pictures or scenarios evolve, depending on the experimental methods used and on the momentum scale under consideration:

- *The quark and gluon view in the perturbative regime* emerged from the deep-inelastic scattering of leptons on hadrons. For high-energy particle momenta, asymptotic freedom governs the behavior of the particles, and their properties can be observed in great detail. In this case, the measured properties are directly related to the quark and gluon degrees of freedom as they appear in the QCD Langrangian.
- *The hadron view, which is dominated by non-perturbative effects,* appears to be the natural perspective to take when studying quark objects in their rest frame or at lower energies. At these energies, where quarks seem to be "dressed" with gluons and quark-antiquark pairs, the connection of the QCD Lagrangian to the microscopic description is inadequate because the individual quarks and gluons making up the many-body system of the proton cannot be removed and examined in isolation. Thus it makes sense that for hadrons, the degrees of freedom of the QCD Lagrangian are not the degrees of freedom that are important in describing properties like the charge radius, magnetic moments, or the results of spectroscopy experiments in general. Phenomenological quark models often give a better description of experimentally observed results than would be expected. Yet the quantitative description of hadron properties in terms of quarks and gluons remains a challenge. Further progress will require high-precision data input from experiments like $\overline{\text{P}}\text{ANDA}$.

Spectroscopy experiments were at the heart of physics in the past century. Their results led, for example, to the development of quantum mechanics and the quark model. However, if we could really understand the strong interaction by QCD, hadron spectroscopy nowadays would be a dull rather than a challenging enterprise. In fact, the contrary seems to be the case: as the work has gone to deeper levels, the open problems have increased. Even the heavy-quark states that were thought to be well understood have continued to produce many surprises. This shows that the understanding of long-distance dynamics is still somewhat primitive and we still have a great deal to learn from future spectroscopy experiments. In particular, new forms of hadronic matter like multiquark states, glueballs or hybrids will deepen our understanding of the strong interaction.

Clearly missing so far is the unambiguous identification and detailed study of gluonic hadrons. The self-interaction of gluons that is central to QCD leads to a flux tube of gluons exchanged between the quarks that are bound by them. In consequence, quarks cannot be separated infinitely and are subject to confinement. It is a fact that individual free quarks have never been observed experimentally to date. The self-interaction in phenomenological models and in lattice gauge theory calculations clearly points to the existence of bound states of pure glue, known as glueballs. Furthermore, states in which excitations of the gluonic flux between the quarks contribute to the overall quantum numbers of a q$\bar{\text{q}}$ pair should exist; these are called hybrids. In the ideal case for experimental observation, the quantum numbers do not follow the rules

$$\vec{J} = \vec{L} + \vec{S}$$
$$P = (-1)^{L+1}$$
$$C = (-1)^{L+S}$$

that are required for a fermion-antifermion pair in a meson. S denotes the total spin while L stands for the angular momentum between the two fermions. Such exotic quantum numbers ($0^{--}, 0^{+-}, 1^{-+}, 2^{+-}, 3^{-+}, ...$) are an unambiguous sign of a state that cannot be accommodated within the NQM.

The second generation of hadron physics spectroscopy experiments in the 1980's and 1990's were designed to look especially for gluonic excitations. The experiments were mostly located at CERN, but the electron-positron colliders at Beijing and Cornell also played a role.

From the theoretical point of view, it seems reasonable that the antiproton-proton annihilation process is a gluon-rich environment because the antiquarks that are initially present annihilate into gluons. A similar mechanism is present in the radiative decay of heavy quarkonia into light quarks. In the case of the J/ψ, the photon carries away the spin while the gluons from the antiquark-quark annihilation process form a glueball. The central production of a meson in a pp collision should also be a gluon-rich process. This is believed to proceed via double-Pomeron collision, where the Pomeron itself is a multi-gluon state.

At CERN, a unique facility came into operation in 1983 with the low-energy antiproton ring (LEAR). Until its closure at the end of 1996, LEAR provided pure and high-intensity antiproton beams (up to $2 \times 10^6$ $\bar{\text{p}}$/s) in the momentum range between 60 and 1940 MeV/c with a small momentum spread of $\Delta p/p = 10^{-3}$. A particularly interesting experimental aspect was that the new experiments at LEAR could identify charged and neutral particles in a 4π arrangement and had special trigger capabilities, e.g., on photons (Crystal Barrel [3, 5]) or on charged kaons (OBELIX [4, 5]).

The Crystal Barrel experiment and the OBELIX experiment focused on antinucleon-nucleon annihilations at an extracted beam. The JetSet [5] experiment at CERN played a special role since it was located inside the LEAR ring and was running without a magnetic field. The outstanding feature of Crystal Barrel was an electromagnetic crystal calorimeter of CsI crystals for the high-resolution measurements of photons produced in the annihilation process. Since no previous hadron-spectroscopy experiment utilizing antiprotons had had a 4π electromagnetic crystal calorimeter combined with charged tracking as Crystal Barrel did, this marked the beginning of the time that experimental physicists could finally investigate the 60% of annihilation channels that contain more than one neutral particle in the final state. OBELIX's particular strength lay in identifying charged kaons. Both detectors, Crystal Barrel and OBELIX, were thus the first ones in hadron physics to detect both charged and neutral particles with high accuracy complemented by high statistics. This proved to be the key in the Dalitz plot analysis and subsequent partial-wave analysis (PWA) to determine the quantum numbers of the particles produced.

The WA 102 experiment at CERN used the collisions of protons with a proton target to study hadrons in a central production experiment. In central production, the total energy is shared between the meson states produced in the Pomeron-Pomeron collision and the recoiling protons. The creation of new particles is limited by the center-of-mass energy of the Pomeron-Pomeron reaction. Since it is impossible to determine the mass and quantum numbers of the produced particles from the initial state, the analysis is often rather complicated or even ambiguous. For the WA102 experiment, which used a 450 GeV/c proton beam onto a fixed target it turned out to be difficult to look for mesons in the charmonium region due to energy limitations [6].

At electron-positron colliders, heavy quarkonia with the quantum numbers $J^{PC} = 1^{--}$, like the $\Upsilon$ or the $J/\psi$, can easily be used to look for subsequent radiative decays. $\Upsilon$ decays were investigated by the CLEO collaboration at Cornell [7], but suffered severe drawbacks from several suppression factors as compared to radiative $J/\psi$ decays. Especially the radiative decays of $\Upsilon$ resonances are naively expected to be almost an order of magnitude lower than for radiative $J/\psi$ decays. The latter are being studied by the BES collaboration in Beijing, but the BES II detector had a weakness in photon detection due to the lack of a crystal calorimeter [8]. Nevertheless, the information gathered in the different experiments still complement each other rather well.

**Gluonic excitations**

Glueballs pose a very interesting facet of nature. The gluons are the mediator particles of the strong interaction and themselves massless. However, they do carry color, i.e., the charges of the strong interaction. It is generally believed that there is an attractive interaction between gluons leading for example in lattice calculations based on the QCD Lagrangian to a flux tube of gluons between the quark and the antiquark in a meson. The mutual gluon attraction should also allow for the formation of meson-like bound states of gluons, even if no quarks are present. Lattice calculations predict a whole spectrum of bound gluon states, the glueballs, which are shown in Figure 1.

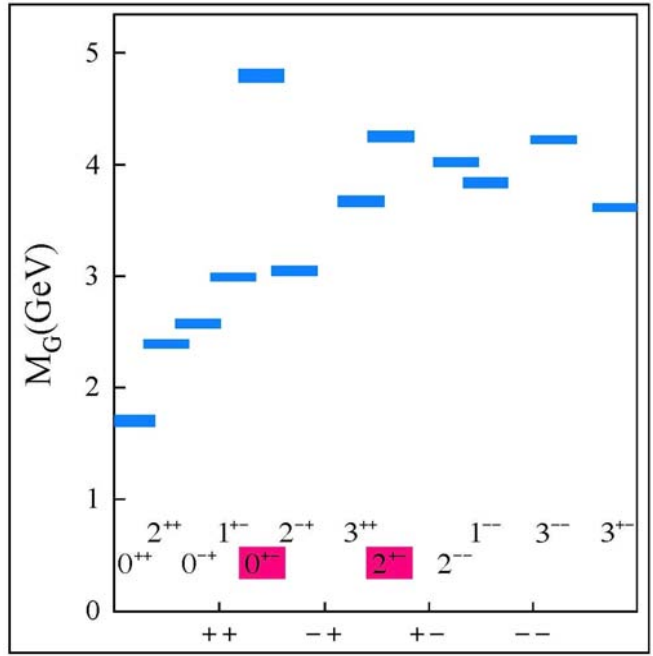

Figure 1: The predicted glueball spectrum in lattice calculations taken from [9]. Exotic quantum numbers are marked with colored boxes.

While the Higgs mechanism might be responsible for the masses of the elementary particles, the mass-creation mechanism for hadrons is quite different. Only a few percent of the mass of the proton is due to the Higgs mechanism. The rest is created, in a to date not very well determined way, by the strong interaction. In particular, glueballs themselves would be massless without the strong interaction and their predicted masses arise solely from the strong interaction. Glueballs thus offer a unique way to study the mass creation of strongly interacting particles.

Already in the *Feynman lectures on Gravitation* there has been mentioning of a hypothetical graviton-graviton interaction. Clearly it is hard to see how a graviton-graviton interaction could be measured in a controlled manner, but glueballs do offer at least the possibility to study the non-perturbative interaction of other massless gauge bosons beside gravitons in a controlled manner in the laboratory. It remains to be seen if the results from glueball studies have implications for gravitational forces.

A good candidate for a quantum theory of gravitation is superstring theory. String theory was originally developed as a theory to attempt to describe hadrons. However, it was quickly realized that a four-dimensional string theory had a lot of problems. To solve those problems, the strings must contain both bosonic and fermionic degrees of freedom, which are connected by supersymmetry. Also, strings make sense only in a 10-dimensional space-time. Both aspects seem incompatible with the structure of the strong interaction. Nevertheless there are tantalizing similarities between superstring theory and the strong interaction. In superstring theory, gravitation is represented by closed strings, and in lattice calculations glueballs are also seen as closed loops of gluons. This picture goes well together with our paper on the structure of glueballs, in which we have proposed closed fluxes of colored gluons with twists or knots that give glueballs their distinct masses and quantum numbers [10]. If indeed a connection between superstrings and certain aspects of the strong interaction exists, a detailed study of glueballs might be very rewarding to understand gravitation better.

In the past years, a new relation between modern superstring theory and QCD has been developed, the so-called AdS/CFT (Anti-de Sitter space/conformal field theory) correspondence. AdS/CFT provides a map between a weakly coupled supergravity theory and a strongly coupled large-N gauge theory [11]. The further development of superstring theory models then inspired phenomenological approaches to QCD, called AdS/QCD, which allow predictions of meson masses and couplings that are typically accurate to 10%. Predictions for glueball masses also exist. For a detailed review of this topic, see the review article written by J. Erdmenger [12].

How can the unknown structure of glueballs and their properties be addressed and studied? The spectroscopy is an approach with proven success and if feasible with several glueballs it could indeed turn out to be the only realistic way to deepen knowledge in this area. Figure 1 shows that the lightest glueball, according to lattice calculations, is the scalar glueball with the quantum numbers $J^{PC} = 0^{++}$. In the mass region it is predicted, however, many mesons are present with the same quantum numbers. Mixing between overlapping resonances might occur so the interpretation as glueball is complicated by the mixing of the glueball with the broad light-quark states.

Experimentally, the situation is the following: the simplest way to look for scalar resonances with $J^{PC} = 0^{++}$ is through antiproton-proton annihilation at rest into three pseudoscalars. In such a process, the scalar resonance decays into two pseudoscalars ($0^+ \rightarrow 0^- \, 0^-$), while the third recoiling pseudoscalar removes the excess energy. No angular momentum barrier is present. Such processes have been investigated in the past, mainly involving charged pions, but the annihilation process into charged pions is dominated by the production of the ρ(770), which complicates the analysis of the underlying scalar resonances.

Before the Crystal Barrel experiment at LEAR came into operation, the channel $\bar{p}p \rightarrow 3\pi^0$ had a data sample containing only 2100 events from optical spark chambers [13]. Crystal Barrel obtained extremely high statistics of approximately 700,000 events for this reaction [14]. The Dalitz plot is shown in Figure 2.

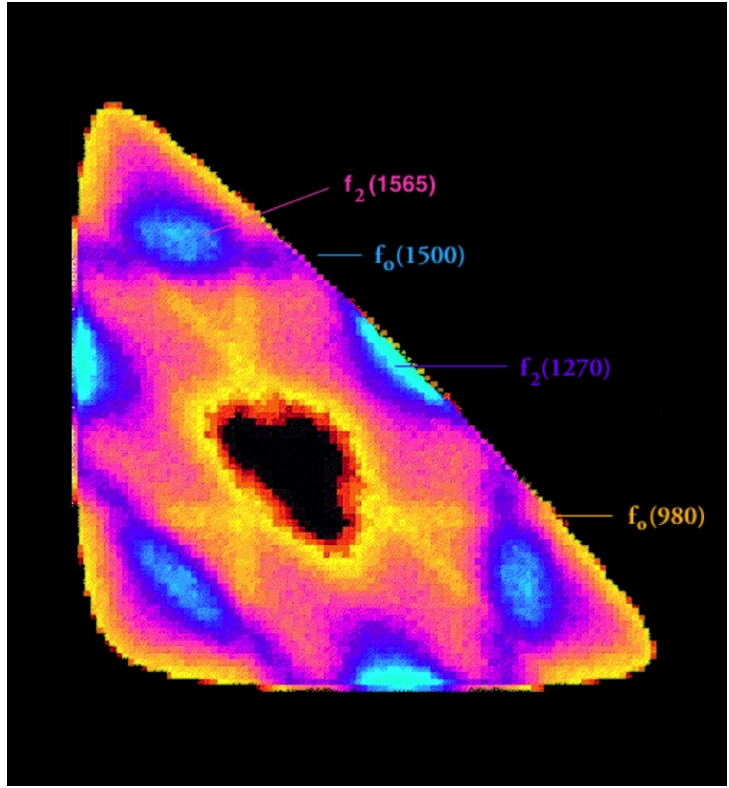

Figure 2: The high-statistics $3\pi^0$ Dalitz plot containing 700,000 events, as measured in $\bar{p}p$ annihilations at rest by Crystal Barrel at LEAR. Each event appears 6 times for symmetry reasons.

Besides the known resonances $f_0(980)$, $f_2(1270)$ and $f_2(1565)$, the $3\pi^0$ Dalitz plot shows two new resonances: $f_0(1370)$ and $f_0(1500)$.

The $f_0(1500)$ has also been seen by Crystal Barrel in its decay modes into $4\pi^0$ ($\bar{p}p \to 5\pi^0$), $\eta\eta$ ($\bar{p}p \to \eta\eta\pi^0$), $\eta\eta'$ ($\bar{p}p \to \eta\eta'\pi^0$), and $K_L K_L$ ($\bar{p}p \to K_L K_L \pi^0$). An analysis of these channels, partly as a coupled-channel analysis using the K-matrix formalism [15, 16, 17], further constrains the masses and widths of these resonances:

$$f_0(1370): M = 1360 \pm 23 \text{ MeV}, \; \Gamma = 351 \pm 41 \text{ MeV}$$
$$f_0(1500): M = 1505 \pm 9 \text{ MeV}, \; \Gamma = 111 \pm 12 \text{ MeV}$$

It is noticeable that despite its mixing, the $f_0(1500)$ is rather narrow compared to other scalar mesons.

The high statistics of the Dalitz plot turned out to be the key for finding the $f_0(1500)$. This is demonstrated in Figure 3, where the same Dalitz plot is shown with different statistics. It is obvious to the naked eye that the statistics for several thousand events is insufficient to unambiguously claim the discovery and the existence of the $f_0(1500)$.

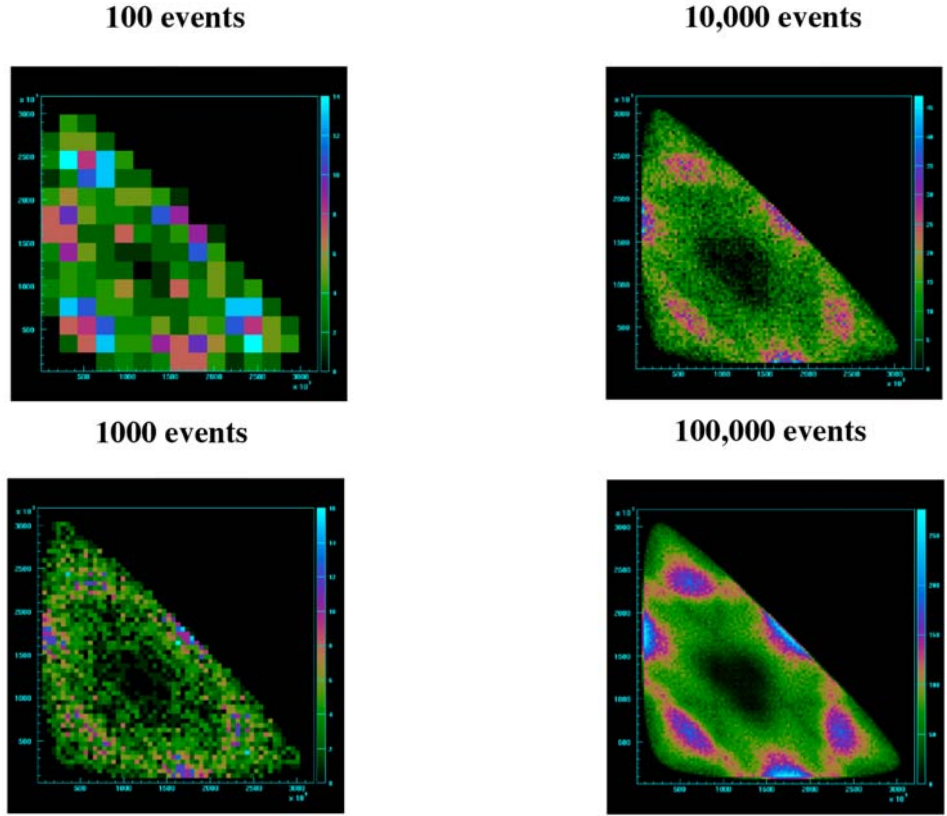

Figure 3: The 3 $\pi^0$ Dalitz plot with different statistics. When the number of events drops to 10,000 the structures start to become ambiguous.

While the experimental existence of the $f_0(1500)$ is beyond any doubt and has been seen by many experiments since, its theoretical interpretation presents a more difficult task. The lowest-mass nonet of scalar mesons could contain the following resonances: $K_0^*(1430)$, $a_0(1450)$ and two open places for which three good candidates exist: $f_0(1370)$, $f_0(1500)$ and $f_0(1710)$. These three states overlap and mix with each other, and since the quark model can accommodate only two of them, the supernumerous one is likely to be a glueball [18, 19].

Originally, the belief was that a glueball could be clearly identified by its flavor-blind decay pattern, i.e., gluons "should not care" about the flavor of the quarks into which they decay. The most obvious glueball candidates among the scalar particles are the $f_0(1500)$ and the $f_0(1710)$. Both of them fit the lattice predictions for a scalar glueball; however, neither of them has a flavor-blind decay. This is attributed to the mixing of all three states. It is remarkable that despite the fact that they are heavily mixed, both glueball candidates have relatively narrow widths of 109 MeV and 135 MeV. This might be a hint that in the absence of mixing, glueballs actually are relatively narrow states.

Since the flavor-blind decay mode does not exist for the $f_0(1500)$, are there other reasons to suspect it might be a glueball? γγ collisions are supposed to act as an anti-glueball filter since photons do not couple to uncharged gluons. The ALEPH collaboration has searched, in γγ collisions, for an $f_0(1500)$ decay into ππ and found no evidence for it there [20].

Could mixing, which destroys the flavor-blind decay mode, be avoided or at least made less probable? Glueballs with exotic quantum numbers (i.e., those that are forbidden for a fermion-antifermion pair such as an antiquark-quark pair in a meson) would suffer less from mixing since resonances with exotic quantum numbers are expected to be rare. Two of the glueballs shown in Figure 1 have exotic quantum numbers; they lie between 4 and 5 GeV.

The same lattice calculations that produced Figure 1 also predict that the majority of glueballs will lie in the mass range between 3 to 5 GeV, i.e., overlapping with the mass region of charmonia. In this mass region, the meson states are much less numerous than in the light-quark sector and seem to be rather narrow, which automatically reduces the problem of mixing. The assumption of relatively narrow states in the charmonium-mass region seems to be confirmed by recent discoveries of additional narrow states (for details, see the chapter "Spectroscopy of X, Y, Z states," below).

One of the newly discovered states is especially interesting in the context of glueball searches: the Y(4140), which was discovered by the CDF collaboration in the decay of B mesons. In the $B^+ \to J/\psi \phi K^+$ decay, evidence (3.8σ) for a narrow J/ψϕ structure was reported in 2009 [21]. The signal was based on an integrated luminosity of 2.7 fb$^{-1}$. Further analysis with higher statistics (6.0 fb$^{-1}$) confirmed the early result (see Figure 4) [22], which now has a significance of more than 5σ.

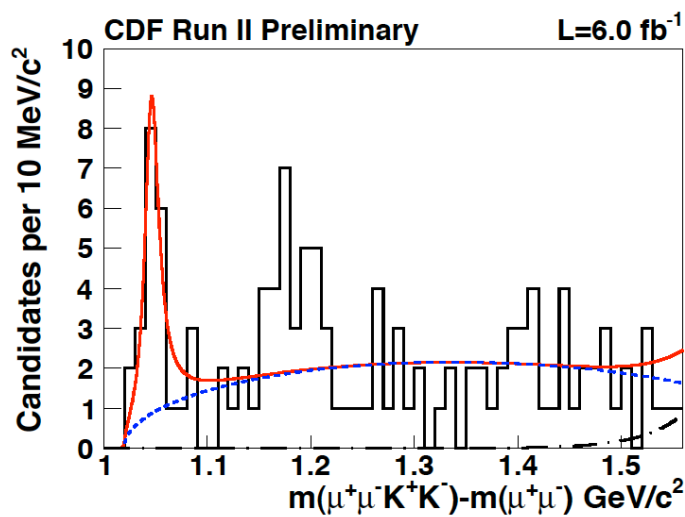

Figure 4: The Y(4140) signal observed by CDF in high-energy $\bar{p}p$ annihilations at Fermilab [22].

The mass and width of the Y(4140) are:

$$M = 4143.4^{+2.9}_{-3.0} \text{ (stat)} \pm 0.6 \text{ (syst) MeV}$$

$$\Gamma = 15.3^{+10.4}_{-6.1} \text{ (stat)} \pm 2.5 \text{ (syst) MeV}$$

The decay width of 15.3 MeV points to a strong rather than an electromagnetic or weak decay, but is nevertheless much narrower than most strong decays—this is reminiscent of the situation with the $f_0(1500)$ and $f_0(1710)$ in the light-quark sector. It was pointed out by the CDF collaboration that the Y(4140) lies well above the open charm threshold; so if it were a normal charmonium state, it would decay predominantly into open charm pairs [21]. What is very intriguing with this state is that it is the first particle found to decay into two heavy quarkonia ($c\bar{c}$ and $s\bar{s}$), which looks exactly like a flavor-blind decay as is expected for glueballs. The Y(4140) is unlikely to be a $J/\psi\phi$ resonance since no such resonance is predicted to exist, and in mass it seems too far away from the threshold to be considered a molecule. What speaks against the Y(4140) being a glueball is that B decays are generally not considered to be gluon-rich. On the other hand, it is certainly worth noting that the Belle collaboration has searched for the Y(4140) in $\gamma\gamma$ collisions but has not found it [23], and this non-detection is what one would expect from a glueball candidate. In order to make progress in understanding the Y(4140), it would be necessary to establish other possible decay modes and determine its properties in detail. Since for statistical reasons this is possible neither in CDF nor at the B factories, only the $\overline{P}ANDA$ experiment will have the possibility to do so in the foreseeable future.

$\overline{P}ANDA$ should also be able to measure all glueballs below 5.4 GeV. The gluon-rich annihilation process has no restrictions on the quantum numbers of states that can be reached; this is in contrast to $e^+e^-$ colliders, where only $J^{PC} = 1^{--}$ can be produced and where all other states are seen solely in the decay chain of the vector state. Thus $\overline{P}ANDA$ will be the only experiment to be able to do systematic studies of supposed glueball states, revealing their true nature and hopefully also giving insights into their structure.

Primary channels for identifying glueball decays are decay channels into heavy quark pairs that would otherwise be rare; thus, the decay into a charmonium state and an $s\bar{s}$ meson would provide an excellent trigger, if allowed by conservation laws, but studies of decay channels into $\phi\phi$ are potentially also very rewarding. The requirements on the $\overline{P}ANDA$ detector to detect such decays are (1) for charmonium decays, good lepton-pair identification by means of an electromagnetic calorimeter or by means of muon detectors and (2) for $\phi$ decays, a particle-identification system that recognizes kaons. Alternatively, because the $\phi$ can also decay into $K_S K_L$, a micro vertex detector—which can detect a multiplicity increase of charged particles due to the $K_S$ decay into two charged pions—is an extremely helpful tool in triggering.

Performance in detecting glueball decays into $\phi\phi$ with $\overline{P}ANDA$ via Dalitz plot analyses was studied on four glueball candidates, all of which were allowed to decay into $\phi\phi$ and produced in association with either a $\pi^0$ or $\eta$ meson [24]. The mass values and spin assignments were taken from lattice calculations [25] and are given in Table 1.

Table 1: Predicted glueball states that are allowed to decay into ϕϕ

| $J^{PC}$ | $M_{glueball}$ |
|---|---|
| $2^{++}$ | 2390 (30) (120) |
| $0^{-+}$ | 2560 (35) (120) |
| $2^{-+}$ | 3040 (40) (150) |
| $3^{++}$ | 3670 (50) (180) |

The assumed width for the glueballs in the simulation was 10 MeV. The reconstruction efficiency for $\overline{P}$ANDA depends on the selection criteria. Since the final state contains kaons, particle identification is a powerful tool for triggering and analysis. In the offline analysis, the detector response is used to determine the probability that a particle is a kaon instead of a pion. For any given event, we can require a higher or a lower probability for the particle to be a kaon, which then influences the overall reconstruction efficiency. Figure 5 shows the reconstruction efficiency for $\overline{P}$ANDA dependent on the momentum of the incident antiproton. Healthy efficiencies of between 5 and 19% can be expected for the different kaon identification probabilities.

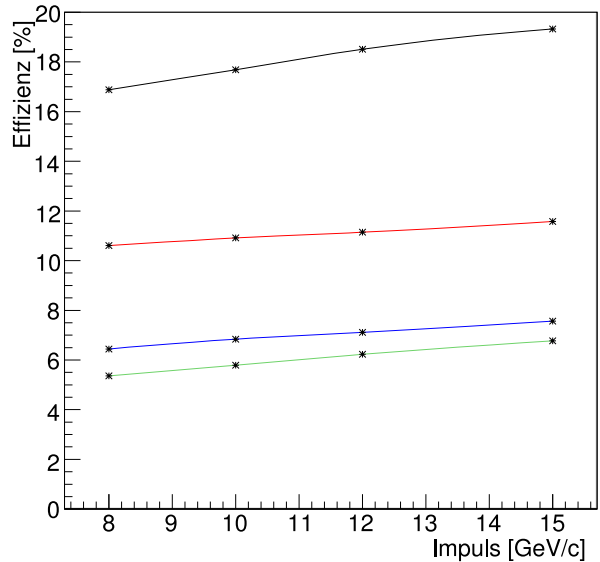

Figure 5: Reconstruction efficiencies for final state G(3670)η for different antiproton momenta and particle-identification criteria of varying harshness shown as different colored curves.

For the Monte Carlo simulation, the assumed cross section for the glueball production was estimated to be 1 nb. This is comparable to the production cross section of normal mesons in this mass region. This assumption for cross sections for QCD exotics were based on the experience with exotic particles gathered by LEAR and Fermilab experiments, where exotic particles seemed to be produced with similar cross sections as ordinary mesons. The total background cross section was estimated from previous measurements and adjusted to the appropriate energy with a Monte Carlo event generator based on the dual parton model (DPM) [26]. The main contributors to the background cross section are pion channels (namely $\pi^+\pi^-\ \pi^+\pi^-\pi^0$, $\pi^+\pi^-\ \pi^+\pi^-\eta$ and $\pi^+\pi^-\ \pi^+\pi^-\pi^0\ \pi^0$), adding up to a cross section of 51 mb that is several orders of magnitude bigger than the signal cross section. However, according to the Monte Carlo studies, the identification of glueballs should nevertheless be possible with an excellent signal-to-background ratio of 10:1. The reconstructed invariant mass of the G(3670) glueball produced with an antiproton beam with a momentum of 15 GeV/c is shown in Figure 6.

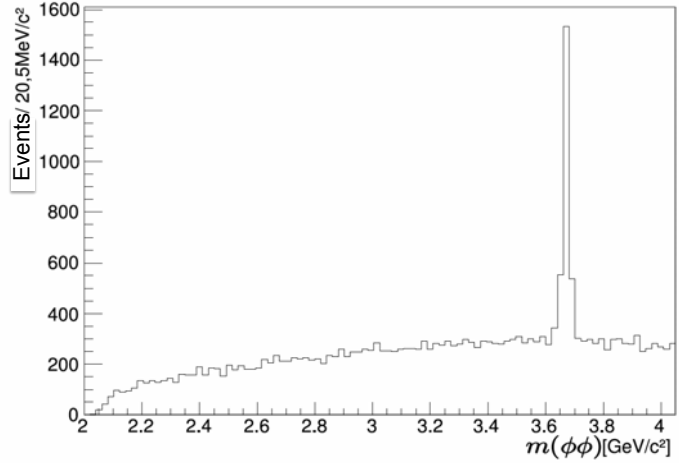

Figure 6: The reconstructed $\phi\phi$ invariant mass in the reaction $\bar{p}p \to G(3670)\eta$ at 15 GeV/c antiproton momentum. Phase-space-distributed events are shown as background.

Table 2 gives the lowest cross section for the four glueballs studied for which a signal-to-background ratio of 10:1 is still achievable in $\overline{P}ANDA$.

Table 2: Lowest values for the cross section $\sigma_S \cdot BF(G \to \phi\phi)$ for which a signal to background ratio of 10:1 can be achieved in $\overline{P}ANDA$. The glueball is produced at an antiproton momentum of 15 GeV/c and recoils against a $\pi^0$ or a $\eta$

| $m(f^{GB})$ | $\sigma_S \cdot BF\left(f^{GB} \to \phi\phi\right)[nb]$ | |
|---|---|---|
| MeV | $\pi^0$ | $\eta$ |
| 2390 | 1.48 | 3.67 |
| 2560 | 1.52 | 3.67 |
| 3046 | 1.36 | 3.52 |
| 3670 | 1.24 | 3.27 |

When it comes to exotic glueballs, the annihilation of antiprotons with protons has another supreme advantage compared to other reaction mechanisms. While the process of a complicated partial wave analysis usually leaves some ambiguities concerning identifications of an exotic nature, such identifications can be easily established in antiproton-proton annihilations. In a production experiment, where the particle is usually produced together with a light meson, exotic quantum numbers are allowed for the unknown state. A subsequent formation experiment with the center-of-mass energy exactly at the particle mass should then allow the particle's properties to be scanned—unless, of course, it has exotic quantum numbers, in which case it cannot be reached in a formation experiment. Thus the appearance of a resonance in a production experiment and its absence in a subsequent scan shows immediately its exotic nature.

Beside glueballs, hybrids provide another interesting approach to gluonic excitations. In a hybrid, a quark-antiquark pair is bound by a flux of gluons that is in an excited state. It would be important to understand how this gluonic excitation translates into properties and decay modes of particles. The quark-antiquark pair could be composed of either light quarks (u, d or s) or charm quarks.

Early antiproton experiments at LEAR looked for light hybrids. Crystal Barrel, for example, searched for hybrids with the exotic quantum numbers $J^{PC} = 1^{-+}$. They are exotic because a normal $q\bar{q}$ pair is a fermion-antifermion system and cannot produce the quantum numbers $1^{-+}$. A state with exactly these exotic quantum numbers had been reported before Crystal Barrel by the GAMS collaboration at CERN [27] in the reaction $\pi^- p \to \pi^0 \eta n$; but a reanalysis of the data gave ambiguous solutions. Also, in the reaction $\pi^- p \to \pi^- \eta p$ in a 18 GeV pion beam at BNL, an asymmetry in the forward/backward $\pi\eta$ angular distribution was explained by the existence of a $1^{-+}$ state [28]. Crystal Barrel studied the $\eta\pi$ and $\eta'\pi$ final states, since hybrids with $J^{PC} = 1^{-+}$ would decay to $\eta\pi$ and $\eta'\pi$ with a relative P-wave between the two pseudoscalar mesons. Such a state would be an isovector (and cannot be confused with a glueball, which must be an isoscalar state).

Crystal Barrel concentrated on the reaction $\bar{p}d \to \pi^-\pi^0\eta p_{spectator}$, which corresponds to the annihilation on a neutron [29]. The advantage of this reaction is that there are no isoscalar resonances that can contribute to this final state. The Dalitz plot with 52,567 events (Figure 7) contains the $\rho(770)$ and the $a_2(1320)$. The strong presence of events in the $\eta\pi$ mass region around 1300 MeV, which appears only above the $\rho$ band, hints at interferences between the $a_2(1320)$ and some other amplitude. Indeed, the partial-wave analysis requires the inclusion of a resonant $\eta\pi$ amplitude with a relative P-wave between the two mesons. The mass and width of this $1^{-+}$ resonance, called $\pi_1$, are:

$$M = 1400 \pm 30 \text{ MeV}$$
$$\Gamma = 310 \pm 70 \text{ MeV}$$

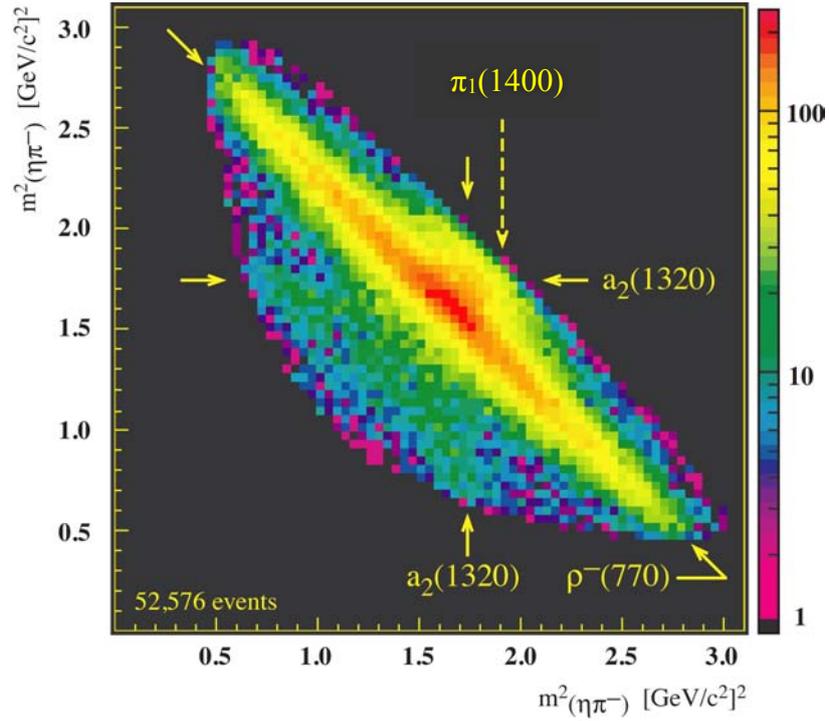

Figure 7: Dalitz plot of the reaction $\bar{p}p \to \pi^-\pi^0\eta p_{spectator}$. The $1^{-+}$ resonance $\pi_1(1400)$ shows up as a bunching of events above the ρ band.

The remarkable outcome, however, is that the $1^{-+}$ resonance takes up 11% of the total intensity of the Dalitz plot, so its production rate in the antiproton-proton annihilation is comparable to those of normal mesons like the $a_2(1320)$, with 15% of the total intensity (see Figure 8).

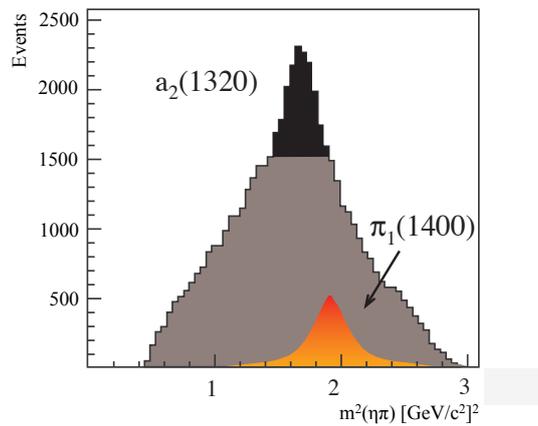

Figure 8: The intensity taken by the $\pi_1(1400)$ and the $a_2(1320)$ in the $\bar{p}p \to \pi^-\pi^0\eta p_{spectator}$ Dalitz plot

The quantum numbers of the $\pi_1(1400)$ are exotic and it could be a hybrid state or a four-quark resonance. However, if it were a four-quark state, one would have expected many more of such four-quarks states to have been reported in other reactions (such as γγ collisions).

In the search for charmonium hybrids, the situation is even more complex. The discovery of many new and unexpected states with unusual properties confuses the situation (see chapter "Spectroscopy of X, Y, Z, states" for details). However, some of the states do decay exactly the way one might expect from a heavy-quark hybrid in a lattice calculation [30]. One could imagine a charmonium hybrid with excited glue decaying into regular charmonium (composed of the relatively heavy charm quarks) and a light meson (from the de-excitation of the gluonic flux). This is good news for $\overline{\text{P}}$ANDA since the leptonic decays of charmonia provide a very clean trigger for events.

As for the lightest charmonium hybrids, where the $q\bar{q}$ pair is in an s-wave configuration, the quantum numbers $1^+$ or $1^-$ are added for a color-electric mode, or $1^-$ for a color-magnetic excitation. In this simple scenario, eight hybrid states should be present, three of them with exotic quantum numbers (Table 3).

Table 3: The quantum numbers of the lowest-lying charmonium hybrid states

| $(q\bar{q})_8$ | Gluon $1^-$(TM) | $1^+$(TE) |
|---|---|---|
| $^1S_0, 0^{-+}$ | $1^{++}$ | $1^{--}$ |
| $^3S_1, 1^{--}$ | $0^{+-}$ ← exotic | $0^{-+}$ |
|  | $1^{+-}$ | $1^{-+}$ ← exotic |
|  | $2^{+-}$ ← exotic | $2^{-+}$ |

The extra gluonic degree of freedom shows up in the confining potential for the hybrid system, for example as derived from Lattice Gauge Theory (LGT) calculations in the Born-Oppenheimer approximation [31]. Figure 9 shows the potentials between the static quarks as a function of their separation distance $R$. The zeroth-order $q\bar{q}$ potential without gluonic excitation, $V_{q\bar{q}}$, applies to the normal charmonium spectrum and describes the usual confining force. The lowest-order hybrid-type potential, $V_{\text{Hybrid}}$, results from the first excited state of gluonic flux and is responsible for the eight lowest-lying hybrid states with different quantum numbers.

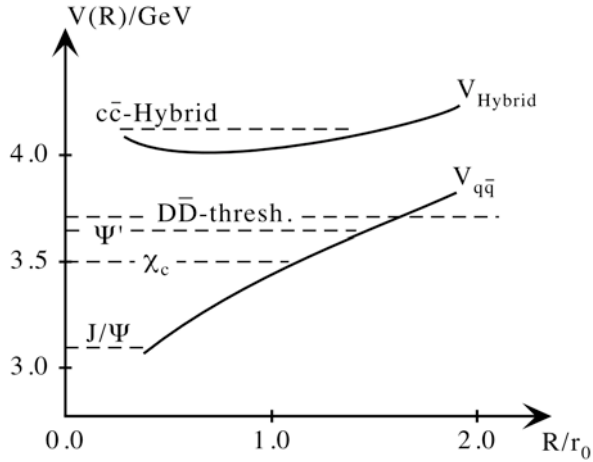

Figure 9: Potentials between static quarks at separation *R*, in units of $r_0 \approx 0.5$ fm, as derived from LGT calculations (scaled from [31]). The $V_{q\bar{q}}$ curve represents the ground-state potential (no gluonic excitation) and corresponds to the conventional charmonium states. . The $V_{Hybrid}$ potential originates from the first excited state of gluonic flux, giving rise to hybrid states.

One of the recently discovered particles that is basically exclusively observed with the unusual decay chain into J/ψ + ω, is the Y(3940). It is interesting that other decay modes, in particular the decay into open charm, have been searched for but turned out not to be important [32]. In studies of the physics performance of the $\overline{P}$ANDA detector, which are the basis of most simulations shown in this report, the possible observation of the Y(3940) was one of the central issues [1]. The cross sections utilized also for background study purposes were estimated, via extrapolation to 3.94 GeV, of the cross sections for the reactions $\bar{p}p \rightarrow \pi^+\pi^-\pi\rho$ and $\bar{p}p \rightarrow \pi^+\pi^-\omega$ that are measured for energies between 2.14 and 3.55 GeV [33]. A comparison was also made to the DPM event generator [26]. The results are shown in Figure 10.

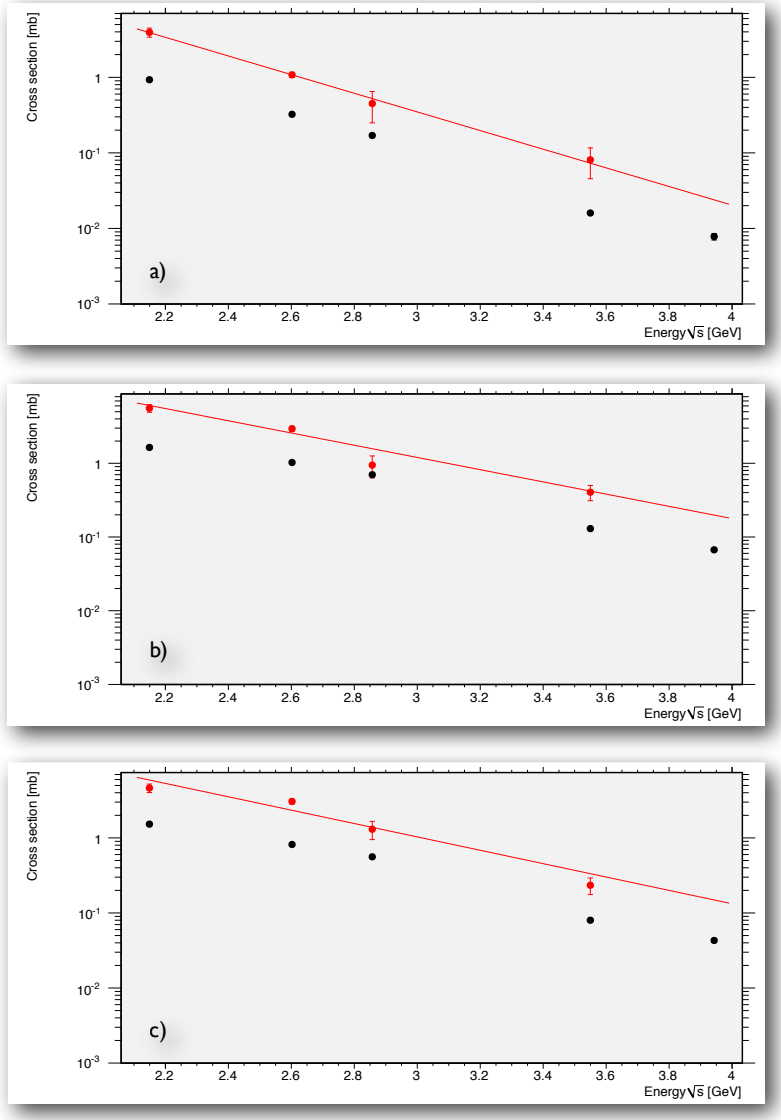

Figure 10: Cross sections for the reactions $\bar{p}p \to \pi^+\pi^-\omega$ (a), $\bar{p}p \to \pi^+\pi^-\pi^-\rho^+$ (b) and $\bar{p}p \to \pi^+\pi^-\pi^0\rho^0$ (c) in dependence of $\sqrt{s}$. The measured cross sections and a fit to the data are shown as solid lines, while points represent the results of the DPM generator.

The assumed cross sections for the signal and possible background reactions are given in Table 4.

Table 4: Cross sections for the signal and background reactions for the detection of the Y(3940). The cross sections marked with * take the branching fraction (BF) of subsequent particle decays into account and are shown as 100%. For the relevant Y decays the appropriate BF is given explicitly.

| Reaction $\bar{p}p \to$ | $\sigma$ | $\mathcal{B}$ |
|---|---|---|
| $Y \to J/\psi\omega$ | $\sigma_S$ | 5.2% x $\mathcal{B}(Y \to J/\psi\omega)$ |
| $\pi^+\pi^-\pi^0\rho^0$ | 149 $\mu b$ * | 100% |
| $\pi^+\pi^-\pi^-\rho^+$ | 198 $\mu b$ * | 100% |
| $\pi^+\pi^-\omega$ | 23.9 $\mu b$ * | 100% |
| $\psi(2S)\pi^0$ | 55 $pb$ | 3.73% |
| $Y \to J/\psi\rho\pi$ | $\sigma$ | 5.9% x $\mathcal{B}(Y \to J/\psi\rho\pi)$ |

After applying the usual data analysis techniques and kinematical fitting, the reconstructed invariant J/ψω mass distribution is shown in Figure 11.

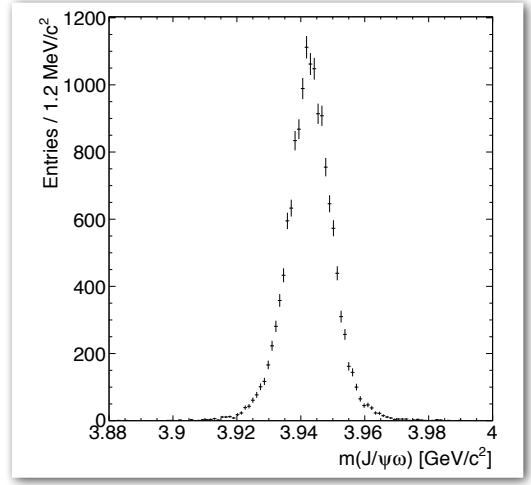

Figure 11: Invariant J/ψω mass reconstructed from $\overline{P}$ANDA Monte Carlo events.

$\overline{P}$ANDA can expect to detect

$$N = \varepsilon_S \tilde{\sigma} = \int \mathcal{L} dt = 66\tilde{\sigma} \ nb^{-1}$$

events per day at peak position with the design luminosity of $\mathcal{L} = 2 \times 10^{32}$ cm$^{-2}$s$^{-1}$ and 50% efficiency. $\tilde{\sigma}$ takes the unknown branching function of Y(3940) → J/ψω into account. The expected signal of several thousand events per day has to be compared with the original signal observed by Belle [34] and BaBar [35] over years of running (Figure 12).

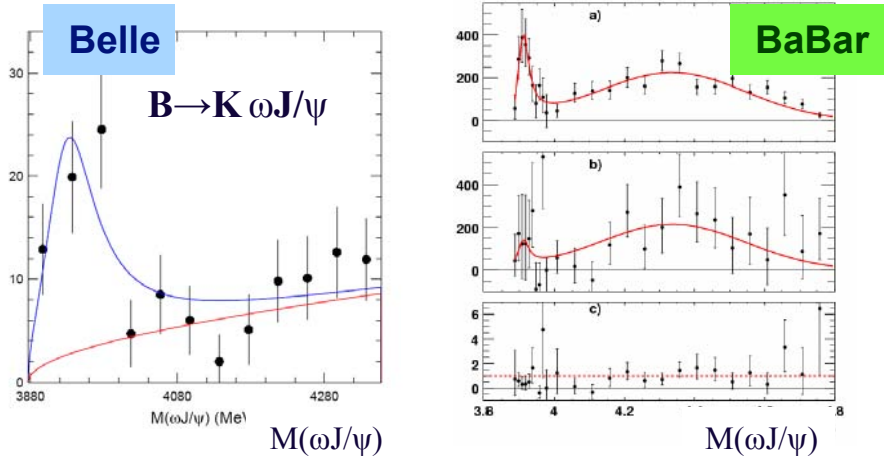

Figure 12: The invariant mass spectrum of the J/ψω system as measured by the Belle collaboration (left) and the BaBar collaboration (right). The picture a) in the BaBar case shows the charged B decays, b) the neutral B decays and c) the ratio between them.

**Spectroscopy of X, Y and Z states**

The initial discovery of the charm quark in the 1970s was followed by the discovery of a quick succession of charmonium states with relatively narrow widths. The narrow widths indicate a long lifetime of the states, which results from the fact that charmonium states below the open-charm decay threshold of 3.73 GeV (i.e., twice the mass of the D-mesons) must decay through the annihilation of the $c\bar{c}$ pair. The spectroscopic investigation of these experimentally clean charmonium states deserves major credit in the development of Quantum Chromodynamics (QCD), the theory of the strong interaction.

The total widths in these charmonium decays have traditionally been described in terms of annihilation into gluons, using the corresponding formulas for positronium annihilation into photons, but with $\alpha_s$ vertices and combinatorial color factors. It was the spectrum of particles consisting of the fourth generation of quarks (charm) and later the fifth generation (bottom) that lent strong support to the viewpoint that mesons could be understood in a way analogous to positronium in the electroweak interactions. In its simplest form, the resulting potential contains a Coulomb-like part based on a one-gluon exchange (OGE) and—special to the strong interaction—a part describing the observed confinement:

$$V(r) = -\frac{4}{3}\frac{\alpha_s}{r} + kr$$

The narrow width of the charmonium states and the fact that they are well separated without interfering amongst each other allowed for precise theoretical calculations to be made. Even the simple-minded OGE complemented with a spin-dependent Breit-Fermi Hamiltonian and a spin-orbit term gave a very good description of the charmonium spectrum as it was known before 2003. Thus the case of charmonium physics seemed to be closed.

From the theoretical side, however, it is questionable whether the concept of "free" gluons is valid in QCD. It seems much more reasonable to describe the binding of the quark-antiquark pair by a flux of multiple gluons confined to a tube due to the gluon self-interaction, as already mentioned earlier.

Interest in charmonium physics got a big boost again in 2003, when the Belle collaboration found an unexpected, very narrow state in the $J/\psi\pi^+\pi^-$ invariant mass spectrum, called X(3872). The latest spectrum from Belle is shown in Figure 13 [36].

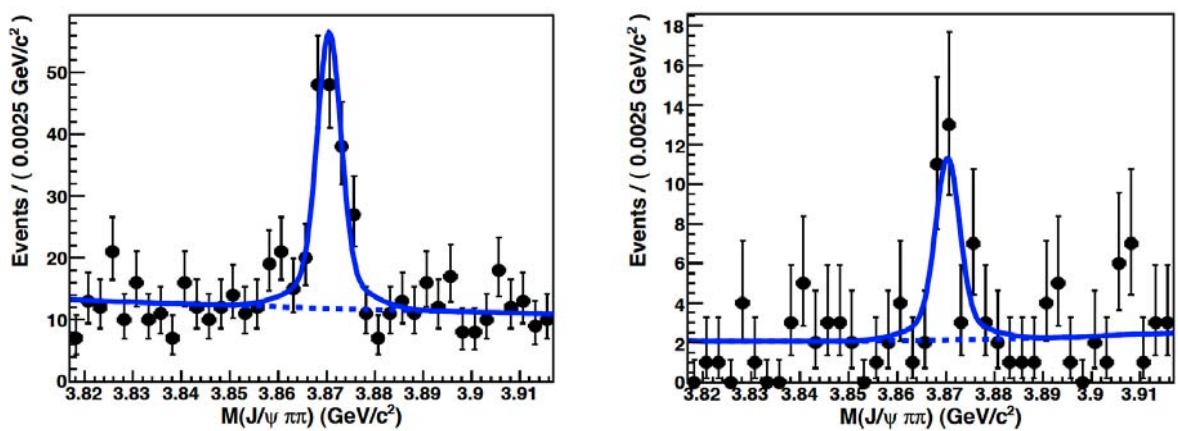

Figure 13: The $J/\psi\pi^+\pi^-$ invariant mass distribution for charged (left) and neutral (right) B decays.

The properties of this state are

$$M = 3871.56 \pm 0.22 \text{ MeV}$$
$$\Gamma < 2.3 \text{ MeV}$$

The Belle collaboration gives as a likely $J^{PC}$ assignment $1^{++}$, making it (together with the narrow width that is so far known only as an upper limit) unlikely to be a conventional charmonium state. What else could it be? It is interesting to recognize that it lies within 0.5 MeV of the $D^{*0}\bar{D}^0$ threshold, and nowadays we know that it decays into $J/\psi\pi^+\pi^-$ and $J/\psi\pi^+\pi^-\pi^0$ with similar rates. The proximity to a threshold makes it a prime candidate for a hadronic molecule. Molecules are weakly bound states of at least two hadrons and nuclei. Here hypernuclei could be considered as well-known examples.

If such hadronic molecules indeed exist, they should appear in the vicinity of a two-particle threshold or between two close-by thresholds, where attractive S-wave interactions can bind a pair of mesons or where a meson and a baryon form this type of resonance. Given this rich set of possibilities, it is one of the mysteries of QCD that so far only quark-antiquark and three-quark states have been firmly established.

Obviously the characteristics for a molecule should be that its mass is slightly less than the mass of the free constituents, due to the binding energy. The interpretation of the state depends critically on the measured values of the state's properties. Various calculations show that depending on what kind of object the X(3872) is, different dispersive effects would result in different line shapes (see Figures 14 and 15, from [37] and [38], respectively).

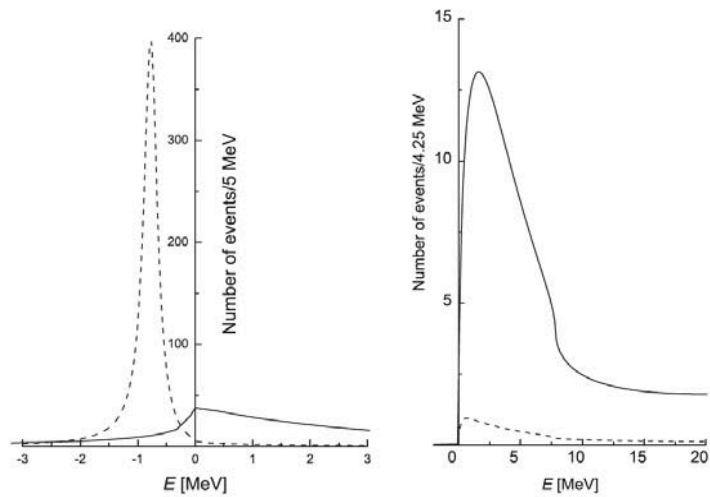

Figure 14: The differential rates for the $\pi^+\pi^- J/\psi$ (first plot) and $D^0\bar{D}^{*0}$ (second plot) for large $\pi^+\pi^- J/\psi$ yield (Belle case) or $D^0\bar{D}^{*0}$ dominance [37]. The Belle fit is shown as solid curve while the model fit is the dashed curve.

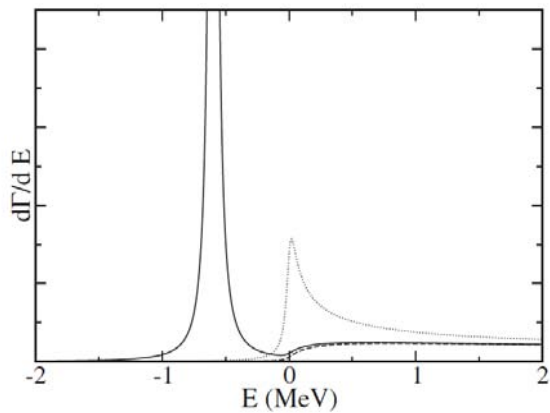

Figure 15: The line shapes near the $D^{*0}\bar{D}^0$ threshold for X(3872) in the $D^0\bar{D}^0\pi^0$ channel. The line shapes correspond to a bound state (solid line), a virtual state (dashed line) and a smooth excitation (dotted line) [38].

The measurement of B decays from BaBar is significantly worse and does not help the situation. The CDF collaboration studied the X(3872) in J/ψπ$^+$π$^-$ decays. They observed approximately 6000 X(3872) events in antiproton-proton collisions at $\sqrt{s}$ = 1.96 TeV [39] (see Figure 16), showing a remarkably strong coupling of this state to $\bar{p}p$. The D0 collaboration confirms the result of CDF [40].

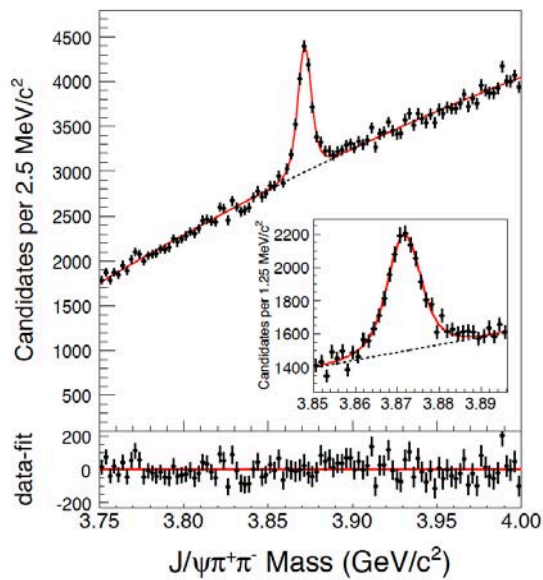

Figure 16: Invariant mass distribution of the X(3872) candidates. The points show the data distribution, the full line is the projection of the unbinned maximum-likelihood fit, and the dashed line corresponds to the background part of the fit. The inset shows an enlargement of the region around the X(3872) peak. Residuals of the data with respect to the fit are displayed below the mass plot.

The mass value of M = 3871.61 ± 0.16 (stat) ± 0.19 (syst) MeV brings the world average of the X(3872) mass within a mass difference of -0.35 ± 0.41 MeV to the sum of masses of the D$^0$ and D$^{*0}$ [41]. Unfortunately the value for the width given by Belle is just an upper limit and has little chance of being improved upon in the immediate future because it is given by the detector resolution; thus a precise determination of the line shape is unthinkable at present.

$\overline{P}$ANDA will change the data situation dramatically. It will scan the parameters of the resonances in formation mode, making use of the excellent momentum resolution Δp/p=10$^{-4}$-10$^{-5}$ of the HESR beam. This technique has proven to be very successful in the Fermilab experiment E835 inside the Fermilab accumulator. Since the parameters of a resonance show up as a formation rate as a function of the center-of-mass energy, the quality of the measurement depends not on the energy resolution of the detector but on the quality of the beam, which is much easier to control. To shown an example, E835 carried out the most precise measurement of the $\chi_{c1}$ by measuring its formation rate through the counting of the subsequent J/$\psi\gamma$ rate [42]. The resulting resonance spectrum can be seen in Figure 17. The uncertainty on the measured values for the mass M = 3510.719 ± 0:051 ± 0:019 MeV and total width $\Gamma$ = 0.876 ± 0.045 ± 0.026 MeV are truly impressive.

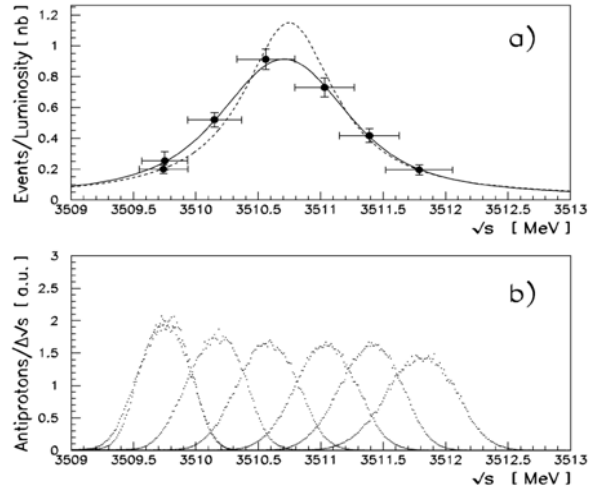

Figure 17: The formation cross section for the $\chi_{c1}$ in $\bar{p}p$ annihilation is shown in a). The horizontal error bar arises from the error in the beam momentum. The spread of the beam momentum itself is shown in b).

One particle that is definitely not a charmonium state or a traditional meson is the $Z^+$(4430). This was discovered by Belle in the K$\pi^{\pm}\psi$' channel (see fig. 18) [43]. This state carries electric charge and thus cannot be a charmonium state even though it must contain a $c\bar{c}$ pair due to its decay into $\psi'\pi^+$. It is an obvious candidate for a multiquark state. Two more things are remarkable about the $Z^+$(4430): its decay into J/$\psi\pi^+$ is not observed, which might be indicative of a certain selection rule, and the width is again rather narrow. It is exactly this narrow width that allowed Belle to rebuke the claim made by the BaBar collaboration that the signal is caused not by a resonance but by interference effects in the K$\pi$ channel. However, the presence of possible interference effects in the Dalitz plot cannot be excluded in B decays, where the K* are naturally present. To finally resolve the issue, one has to go to different channels.

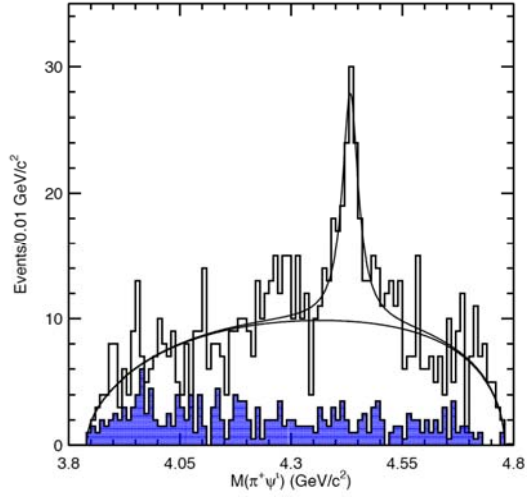

Fig. 18: The $\pi^{\pm}\psi$' invariant mass spectrum for B → K $\pi^{\pm}\psi$' decays. The fit shows a Breit-Wigner and a phasespace-like background. The blue area shows the estimated background.

$\overline{P}$ANDA can study the $Z^+(4430)$ in both production and formation experiments. In the production experiment, the $Z^+(4430)$ would be produced, e.g., in the reaction $\overline{p}p \rightarrow Z^+(4430) + \pi^-$. The subsequent decay chain is then:

$$Z^+(4430) \rightarrow \psi(2S)\pi^+ \rightarrow J/\psi\pi^+\pi^-\pi^+ \rightarrow e^+e^-\pi^+\pi^-\pi^+$$

The reconstruction efficiency for this channel with the $\overline{P}$ANDA detector has been studied in Monte Carlo calculations and is 24%.

Furthermore, the Monte Carlo study shows that the Dalitz plot for the above-mentioned channel is clean and almost background-free (Figure 19) due to the presence and effective detection of the J/ψ in the decay chain. Two high-energy electrons can be clearly identified and distinguished from pions by their shower in the EMC, and together the electrons must have a very narrow invariant mass; both conditions are extremely hard to mimic in any background reaction.

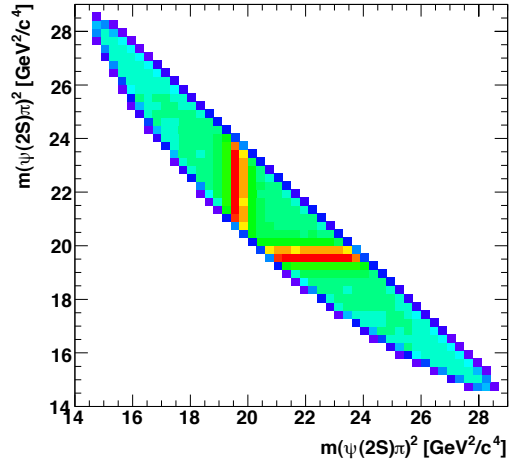

Figure 19: The ψ(2S)π Dalitz plot for the reaction p̄p → Z^±(4430)+π^∓ with the Z decaying into ψ(2S)π.

The invariant ψ(2S)π mass is shown in Figure 20. Since the Z is produced together with an associated pion and there are additional pions from the decay ψ(2S) → J/ψ π^+ π^-, the possibility of a combinatorial background exists. However it turns out that for an incoming antiproton momentum of 15 GeV/c, the "incorrect" combination of pions has a completely different invariant mass distribution.

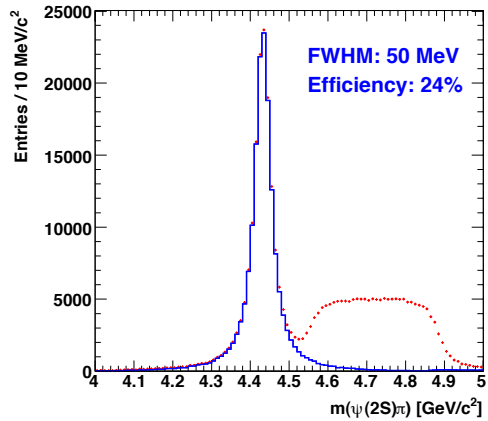

Figure 20: The ψ(2S)π invariant mass spectrum is shown by the blue curve. Combinatorial background from the wrong association of the 3 pions of the final state leads to the distribution in red.

P̄ANDA can investigate the the Z^+(4430) even further by switching to studies of the Z^+(4430) in formation mode. Due to the charge of the Z, this is only possible by annihilating the antiprotons on a neutron in a deuterium target. Experimentally it is no problem to replace the hydrogen gas, for example in a pellet target, with deuterium. The reaction to look for in P̄ANDA would then be:

$$\bar{p}d \to Z^-(4430) + p_{spectator}$$
$$\downarrow$$
$$\psi(2S)\pi^- \to J/\psi\pi^+\pi^-\pi^-$$

The reconstruction efficiency for this channel studied in Monte Carlo reactions is 35% and both the invariant mass of the $Z^+$(4430) and of the $\psi$(2S) can be reconstructed cleanly, as shown in Figure 21.

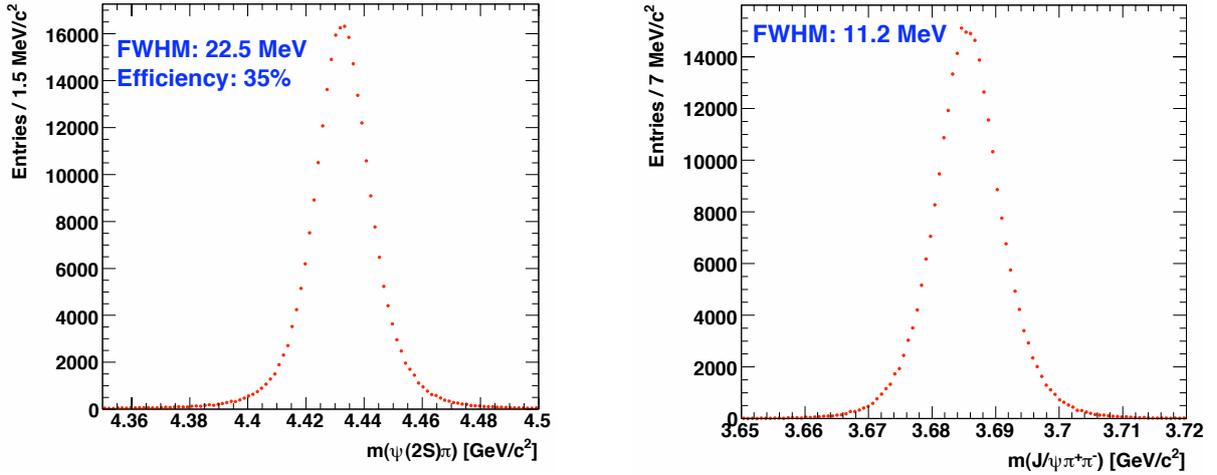

Figure 21: The reconstructed invariant masses of the $Z^+$(4430) and the $\psi$(2S) in $\overline{P}$ANDA Monte Carlo simulations of an antiproton annihilation on the neutron of a deuterium target.

Two more charged states like the $Z^+$(4430) were observed by Belle but not by other experiments—the $Z^+$(4050) and the $Z^+$(4250), both decaying into $\pi^+\chi_{c1}$ [44]. Needless to say, $\overline{P}$ANDA will be able to study both states in a way similar to that described for the $Z^+$(4430).

Many more states, most of them above the open-charm threshold, have since been found by several experiments. (For a review of this topic, refer, e.g., to [45].) Many of the new states do not fit into the expected charmonium spectrum due to their unusual properties or decay patterns. Figure 22 gives an overview of these states in relation to the conventional charmonium spectrum.

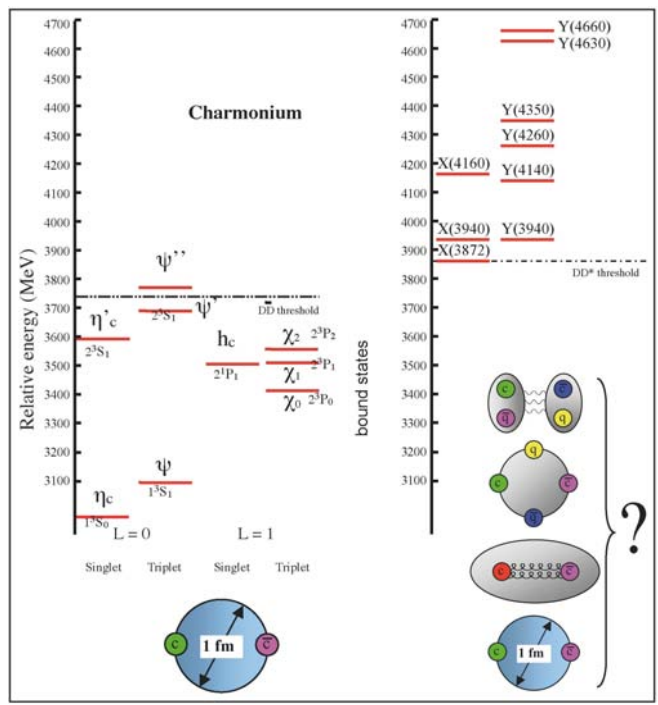

Figure 22: The established charmonium spectrum is plotted on the left side. The right side shows newly discovered X and Y states with unusual properties, whose nature is yet unknown.

Decay patterns that result in a charmonium state and a light meson instead of resulting into open-charm particles raise the question as to whether one or more of the charmonium states could actually be charmonium hybrids. As mentioned earlier, hybrids are quark-antiquark states in which the gluonic degrees of freedom have been excited and contribute to the overall quantum numbers. It seems very plausible that such a gluonic excitation could decay into a light meson, while the relatively heavy charm quarks stay in place and form a charmonium state. A good candidate for a charmonium hybrid is the Y(4260), a state found by BaBar (and subsequently confirmed by Cleo-c and Belle). The Y(4260) decays into $J/\psi \pi^+\pi^-$. The quantum numbers are known from the production mechanism and must be $J^{PC} = 1^{--}$. The $1^{--}$ states in this mass region are well-known from previous $e^+e^-$ studies, and the Y(4260) seems supernumerary. Such overpopulation could also be a further indication for charmonium hybrids.

In order to classify all these new particles and to unveil their nature, their quantum numbers need to be determined and their decay branching fractions—into the different charmonia and possibly open charm—need to be established. For most of these particles, however, the available statistics are insufficient for this (see, e.g., Figure 12), and even less so for measuring the line shape of the resonances to get more insight.

It is therefore worthwhile to turn to the question of how progress could be made in the foreseeable future. In $e^+e^-$ annihilations, direct charmonium formation is possible only for states with the quantum numbers $J^{PC} = 1^{--}$ of the photon, namely the $J/\psi$, $\psi'$ and $\psi(3770)$ resonances. Precise measurements of the masses and widths of these states can be obtained easily by knowing the exact energies of the electron and positron beams. Charmonium states with different quantum numbers can be reached via the decay of these resonances or, at higher energies, by means of other production mechanisms, such as photon-photon fusion, initial state radiation and B-meson decays.

In comparison, antiproton-proton annihilations can access any quantum number and thus all charmonium states, thanks to the multiple gluons present in the process. This was demonstrated by the Fermilab E835 experiment in scanning experiments. The scanning technique became available thanks to the development of stochastic beam cooling. By scanning a resonance in antiproton-proton annihilations, the masses and widths of all charmonium states can be measured with excellent accuracy, since they can be determined through the very precise knowledge of the initial antiproton and proton beams and are therefore not limited by the resolution of the detector. As outlined above, such precision data are necessary to progress further in understanding the conventional charmonium system and the underlying forces. Compared to previous experiments and in order to ensure success, such an antiproton facility should have:

- up to ten times higher instantaneous luminosity (L = $2\times 10^{32}$ cm$^{-2}$s$^{-1}$ in high-luminosity mode, compared to $2\times 10^{31}$ cm$^{-2}$s$^{-1}$ at Fermilab);
- better beam momentum resolution ($\Delta p/p = 10^{-5}$ in high-resolution mode, compared with $10^{-4}$ at Fermilab);
- a better detector (higher angular coverage, the presence of a magnetic field, the ability to detect the hadronic decay modes and to do particle identification).

$\overline{\text{P}}$ANDA and the FAIR antiproton facility were designed keeping these numbers in mind.

What are the cross sections we can expect in $\overline{\text{P}}$ANDA? Very little is known in particular for the production cross sections of charmonia, where the charmonium state is associated with an additional meson in the $\bar{\text{p}}$p annihilation process. Therefore we tried to estimate charmonium production cross sections by using the decay $\psi \to$ m$\bar{\text{p}}$p (where m represents any meson measured), where the results are known and related to the annihilation process by crossing symmetry [46]. This method was refined in a second paper in a hadron pole model [47]. The predicted cross sections vary between 100 pb and 20 nb. Given an integrated luminosity of

$$\int \mathcal{L} dt = 8 \text{ pb}^{-1} \text{ per day}$$

for the $\overline{\text{P}}$ANDA experiment, we can expect between 100 and several thousand charmonia per day in production. The formation cross section is then usually 2-3 orders of magnitude bigger for narrow states. This estimate should also apply to QCD exotics, since we know from previous LEAR experiments and high-energy antiproton collider experiments at Fermilab that exotics are produced with cross sections similar to those of normal mesons.

**Traditional charmonium spectroscopy**

The charmonium spectrum shown on the left side of Figure 22 consists of eight states below the threshold of 3.73 GeV, where the decay into open charm becomes a possibility. All eight states have been established, but their properties have not all been measured with the same accuracy. This reflects the fact that it is easy to form vector mesons in an electron-positron collider, while the singlet states have to be detected in the decay of a triplet state. Only the Fermilab E760 and E835 experiments could achieve the formation of singlet states, but were hampered by the extremely limited beam time available to them. By building upon the experiences of E760 and E835, the physics program of $\overline{\text{P}}$ANDA could contribute significantly to charmonium physics. Some of the open questions will be outlined in the following paragraphs.

*The $\eta_c(2S)$ state* has a mass that corresponds to a surprisingly small hyperfine splitting of $48 \pm 4$ MeV. This hyperfine splitting should be compared to 67 MeV, which is obtained by rescaling the splitting between the J/ψ and the $\eta_c$. The properties of the $\eta_c(2S)$ are not very well measured and need to be improved by $\overline{\text{P}}$ANDA to gain more understanding why the hyperfine splittings are so different.

Precise measurements of the parameters that characterize the $h_c(^1P_1)$ state are needed to resolve a number of open questions in heavy quark physics. The value for the hyperfine splitting of the P states could be used as a sensitive probe of the deviation from a *1/r* potential that is introduced by the spin-orbit force [48]. A measurement of the total width and of the partial width to $\eta_c+\gamma$ will provide an estimate of the decay width to gluons. The calculations of partial gluonic widths of $^{1,3}P_1$ states suffer from infrared divergences to leading order, so a comparison with measured values will be of considerable importance. The ratio of decay branching ratios to two or three gluons in a purely perturbative picture does allow for the determination of the strong coupling constant $\alpha_S$. The triplet states have positive parity and can therefore decay hadronically via two gluons compared to the hadronic decays of the J/ψ and ψ', which can proceed only via at least three gluons. A meaningful comparison between two or three gluon decays requires the measurement of hadronic final states in charmonium decays, which are so far poorly understood theoretically and experimentally and would therefore profit from experimental data taken with $\overline{\text{P}}$ANDA.

*The D states of charmonium* have not been discovered beyond doubt. The J=2 D states ought to be extremely narrow despite being above the open charm threshold. Their annihilation and radiative decays thus provide an interesting test of QCD and related models. In general, charmonium spectroscopy above the $D\overline{D}$ threshold ventures pretty much into "terra incognita" and should therefore be pursued aggressively. That it is very rewarding showed the discovery of the XYZ states.

Even in decays of the well-studied *J/ψ and ψ(2S)* states, there are problems in our understanding. From perturbative QCD, it is expected that the decay of the J/ψ and ψ(2S) into light hadrons is dominated by the annihilation process into three gluons or one virtual photon with a decay width proportional to the wave function at the origin [49]. Therefore the so-called "12% rule" could be formulated:

$$R_h = \frac{BF_{\psi' \to hadrons}}{BF_{J/\psi \to hadrons}} = \frac{BF_{\psi' \to e^+e^-}}{BF_{J/\psi \to e^+e^-}} \approx 12\%$$

In practice, we see huge deviations from this rule in decay channels into ρπ and K*$^+$K$^-$, an inconsistency usually referred to as the "ρπ puzzle." Since the antiproton-proton annihilation process produces several million J/ψ per day, a systematic study of hadronic decays of the J/ψ, the ψ(2S) and the ψ(3S) will be possible with the $\overline{\text{P}}$ANDA detector, which by design is suited for the detection of all hadronic decays.

**Open-charm spectroscopy**

D mesons could be seen as the hydrogen atom of QCD, since a light quark and a heavy quark are bound together. The presence of light quarks introduces aspects of chiral symmetry breaking and restoration; the heavy quark acts in this case as a static source of color. The quark model was capable of describing the spectra of D mesons with reasonable accuracy [50, 51, 52, 53] and even of making predictions, until the new resonances $D_S(2317)$ and $D_S(2460)$ were found in Belle, BaBar and the CLEO experiment [54, 55, 56],

Since the new resonances do not fit well into the quark model, they triggered a whole series of theoretical interpretations. Those range from models that predict the mass splitting of states due to chiral symmetry [57, 58] to those proposing tetraquark states [59, 60, 61, 62, 63, 64, 65, 66] or DK molecules [67].

In order to distinguish between the different models, the precise knowledge of the decay widths of the $D_S$ states certainly would help [68, 69, 70, 71, 72, 73]. The current upper limit of several MeV, given by the detector resolutions of Belle, BaBar and CLEO, is not precise enough to draw any conclusion. $\overline{\text{P}}$ANDA will improve on the width measurment, as explained in the following.

For a narrow resonance, the energy dependence of the production cross section very close to threshold can be calculated in a model-independent way [74]; the function is sensitive to the resonance width. With the beam energy of the HESR storage ring sufficiently well known, a measurement of the energy dependence of the production cross section around the energy threshold should allow the determination of the width of narrow resonances. No requirements are made on the detector resolution. The aim with $\overline{\text{P}}$ANDA is therefore to do a threshold scan of the reaction $\overline{\text{p}}\text{p} \rightarrow \overline{\text{D}}_s\text{D}_{sJ}$ in order to pin down the widths of the $D_{sJ}$ states to within ~100 keV.

The quality of the result in $\overline{\text{P}}$ANDA depends crucially on the signal-to-background ratio and the total beam time invested. As can be see from Figure 23, the fit with a S/B ratio of 1:3 and 14 days at design luminosity reproduces nicely the shape of a resonance which was put into the Monte Carlo calculation with the input values M = 2317.30 MeV and Γ = 1.00 MeV.

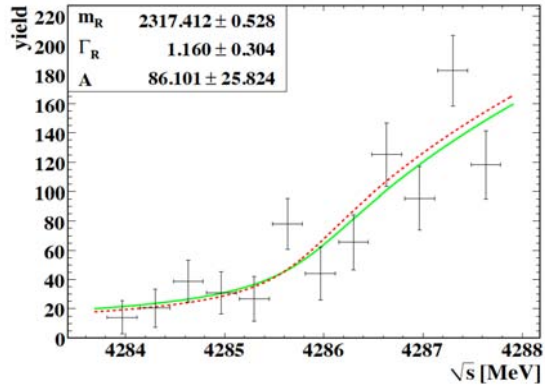

Figure 23: Fit of the excitation function obtained from the reconstructed signal events (green line) and the excitation function corresponding to the generated events (red line).

To see the limits of this method, a similar fit with a S/B ratio of 1:30, 28 days data taking with full luminosity and a width for the resonance of $\Gamma = 0.5$ MeV was performed and did not give meaningful results. The threshold for a successful determination of the $D_{sJ}$ parameters depends crucially on the detector performance of $\overline{\text{P}}\text{ANDA}$, in particular on the identification of secondary vertices with the micro vertex detector and on the kaon identification, which both reduce background signals.

**Baryon spectroscopy**

The baryon spectroscopy of light-quark baryons is pursued intensively at currently existing electron accelerators. Whenever the baryons contain strange or even charm quarks, the data situation becomes extremely sparse. The Review of Particle Physics [75] reports: *"Thus early information about $\Xi$ resonances came entirely from bubble chamber experiments, where the numbers of events are small, and only in the 1980's did electronic experiments make any significant contributions. However, nothing of significance on $\Xi$ resonances has been added since our 1988 edition."* This opens an opportunity for $\overline{\text{P}}\text{ANDA}$ to improve knowledge and observe, for example, the production of excited $S = -2$ and $S = -3$ baryons in the strange-baryon sector. Charmed baryon pair production is also a possibility, but is a question of cross section and has proven to be feasible.

In antiproton-proton annihilations a huge part of the total cross section of 77 mb is associated with a baryon-antibaryon pair in the final state. A particular advantage of using antiprotons in the study of (multi-)strange and charmed baryons is that in antiproton-proton annihilations, no production of extra kaons or D mesons is required for strangeness or charm conservation. The baryons can be produced directly at threshold, which reduces the number of background channels, for example compared to high-energy pp collisions. The production cross section for $\overline{\text{p}}\text{p} \rightarrow \overline{\Xi}\Xi$ has been measured to be 2 μb [76, 77]. Similar cross sections can be expected for excited $\Xi$ states. The triple-strangeness reaction $\overline{\text{p}}\text{p} \rightarrow \overline{\Omega}\Omega$ has never been measured, but theoretical predictions put the price for the production of an additional strange quark pair at a reduced cross section of 2 nb [78]. At design luminosity, about 250 $\bar{\Omega}\Omega$ pairs per day would be produced in the HESR-$\overline{P}$ANDA set-up.

The identification of such a hyperon pair is straightforward if a displaced vertex can be detected. This is illustrated in Figure 24. The key detector element in this case is a micro vertex tracker.

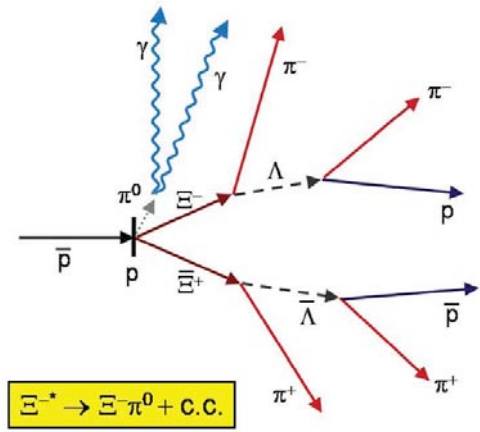

Figure 24: Schematic illustration of the reaction $\bar{p}p \rightarrow \bar{\Xi}\Xi\pi^0$ with subsequent delayed decay vertices.

Monte Carlo studies for $\overline{P}$ANDA give a reasonable and mostly flat reconstruction efficiency of ~15% in $\overline{P}$ANDA over the whole Dalitz plot in the reaction $\bar{p}p \rightarrow \bar{\Xi}\Xi\pi^0$ (Figure 25).

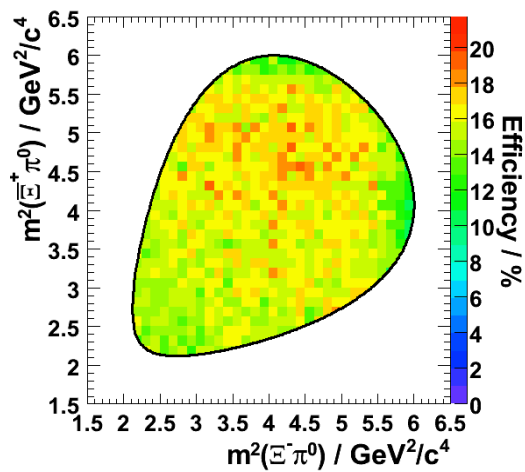

Figure 25: Dalitz plot showing the reconstruction efficiency for $\bar{p}p \rightarrow \bar{\Xi}\Xi\pi^0$

# Hadron Structure Experiments with $\bar{\text{P}}$ANDA

**Previous achievements**

The structure of the nucleon is a defining problem in hadron physics and it can be solved only in the close interplay between theory and experiment. The theoretical understanding achieved there will allow us to get new insights into the underlying theory of strong interaction, QCD.

Hard reactions, such as inclusive deep-inelastic scattering (DIS), semi-inclusive DIS, or hard exclusive reactions are routinely used as probes of hadron structure. The hard scale involved in these processes allows a perturbative QCD description of the reaction, with well-defined operators in terms of quarks or gluons. By determining the matrix elements of such operators in the hadron, the soft part of the amplitude, which lies in the realm of non-perturbative QCD, can be extracted.

The space-like electromagnetic form factors reveal the spatial distribution of the quark charges in a hadron, whereas the structure functions measure the longitudinal momentum distributions of partons in a hadron. However, a full 3-dimensional exploration of the nucleon structure, in both position and momentum space, has only just begun. For this purpose, several correlation functions encoding the quark-gluon structure have emerged in recent years. The correlation between the quark/gluon transverse position in a hadron and its longitudinal momentum is encoded in generalized parton distributions (GPDs) [79, 80, 81, 82]. The information on the quark/gluon intrinsic motion in a hadron and several correlation functions that describe this 3-dimensional structure have been devised in recent years.

**Hard exclusive antiproton-proton annihilation processes**

One of the processes that can be described in certain kinematical domains by the QCD handbag diagram and GPDs is deeply virtual Compton scattering DVCS (see Figure 26). Factorization splits the process into a hard perturbative QCD part and a soft part that can be described by GPDs.

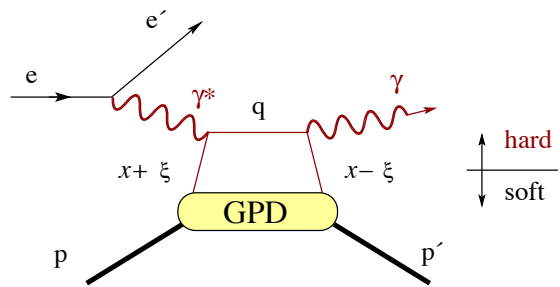

Figure 26: DVCS can be described by the handbag diagram, since there is factorization between the upper "hard" part of the diagram, described by perturbative QCD and QED, and a lower "soft" part described by GPDs.

Another example is the wide-angle Compton scattering process, where the hard scale is related to the large transverse momentum of the final-state photons. The soft part of the process can be parameterized using generalized distribution amplitudes (GDAs). In $\bar{p}p$ annihilations we have the inverted process due to crossed kinematics, as shown in Figure 27.

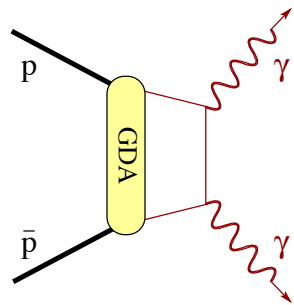

Figure 27: The inverted wide-angle Compton scattering process in $\bar{p}p \to \gamma\gamma$

There are arguments that although it is not accurate for very low or very high energies, the handbag approach might be valid for $\bar{p}p \to \gamma\gamma$ in exactly the energy regime where $\overline{P}ANDA$ operates [83, 84]. Measuring this reaction is a challenge since the two-photon final state suffers from a huge hadronic background. Monte Carlo studies show, however, that due to the excellent electromagnetic calorimeter of $\overline{P}ANDA$, the detection of $\bar{p}p \to \gamma\gamma$ seems feasible despite the background (see Figure 28).

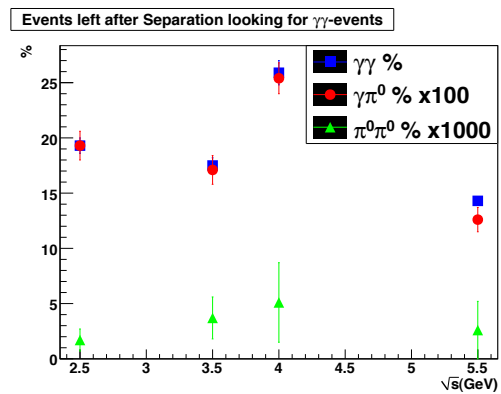

Figure 28: Separation of $\gamma\gamma$ events from background $\pi^0\gamma$ and $\pi^0\pi^0$ events is possible using Monte Carlo corrections. The number of misidentified $\pi^0\gamma$ and $\pi^0\pi^0$ events and their statistical errors are magnified by a factor

100 or 1000 in order to match the limited Monte Carlo statistics to the abundance of the background in the experiment.

**Drell-Yan pair production measurements**

A fast-moving proton can be viewed as a bunch of collinearly moving quarks and gluons. Its inclusive reactions can provide only limited information on the relative motion of these partons. More details of this intrinsic motion are encoded in the transverse-momentum-dependent distribution functions (TMDs). These include spin-dependent correlation functions that link the parton spin to the parent proton spin and to the parton intrinsic motion.

In a Drell-Yan (DY) process, two (polarized or unpolarized) hadrons annihilate into a lepton-antilepton pair. With $\overline{\text{P}}$ANDA only the unpolarized DY process will be accessible, unless the possibility opens up to install a polarized target inside the $\overline{\text{P}}$ANDA detector. The additional availability of polarized antiprotons might be an issue only for later stages of FAIR.

$\overline{\text{P}}$ANDA can determine the Boer-Mulders distribution $h_1^\perp$ by measuring the DY production in antiproton-proton annihilations. In order to obtain such information, the di-lepton mass $M_{ll}$ should be outside the hadronic resonance region, i.e., it should be in the so-called "safe region" between the $\psi'$ and the $\Upsilon$. Since the upper limit of the "safe region" is beyond the energy of the FAIR antiproton complex, $\overline{\text{P}}$ANDA will first concentrate its efforts on the region $1.5 \text{ GeV} < M_{ll} < 2.5 \text{ GeV}$ between the $\phi$ and the J/$\psi$ resonances. There is one advantage of going to this mass region and that is that the cross section in this lower mass region is ~0.8 nb [85], much higher than the 0.4 pb in the traditional "safe region."

**Electromagnetic form factors in the time-like region**

Traditionally, the electromagnetic probe has proven to be an excellent tool for investigating the structure of the nucleon. Compared to the usual lepton-nucleon scattering, in $\overline{\text{P}}$ANDA we are studying the crossed channel process of the antiproton annihilation (Figure 29). This allows $\overline{\text{P}}$ANDA to study certain properties like the time-like form factors of the nucleon with unprecedented accuracy.

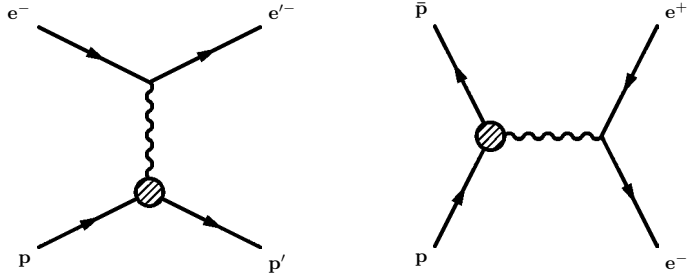

Figure 29: Feynman diagrams for electron-proton scattering and $\overline{\text{p}}$p annihilation

The electric ($G_E$) and magnetic ($G_M$) form factors parameterize the hadronic current and can be determined in both electron-proton scattering and the related crossed process of $\overline{\text{p}}$p annihilation. With electron scattering the form factors can be accessed in the range of negative four-momentum transfer $q^2$ (space-like) while the antiproton-proton annihilation accesses positive $q^2$ values (time-like), starting from the threshold of $q^2 = 4m_p^2$.

In the Breit frame, space-like form factors have concrete interpretations. Since $G_E$ and $G_M$ are Fourier transforms of the spatial charge and the magnetic distribution, their slope at $q^2 = 0$ yields directly the charge and magnetic radii of the nucleon, respectively. In the time-like regime, the form factors reflect the frequency spectrum of the electromagnetic response of the nucleon. By studying both the time-like and space-like regimes, complementary aspects of the nucleon structure are probed and one achieves a complete description of the electromagnetic form factors over the full kinematical range of $q^2$.

The differential cross section for the unpolarized process $\overline{\text{p}}\text{p} \to e^+e^-$ can be written as [86]:

$$\frac{d\sigma}{d\cos\theta} = \frac{\pi\alpha^2(\hbar c)^2}{8m_p^2\sqrt{\tau(\tau-1)}}\left[|G_M|^2\left(1+\cos^2\theta\right) + \frac{|G_E|^2}{\tau}\left(1-\cos^2\theta\right)\right]$$

Measurements over a wide range of $\cos\theta$ permit an independent determination of $|G_E(q^2)|$ and $|G_M(q^2)|$, which has been tried before by other experiments [87, 88] without much success. The world data set on $|G_M|$ under the assumption $|G_E| = |G_M|$ is shown in Figure 30.

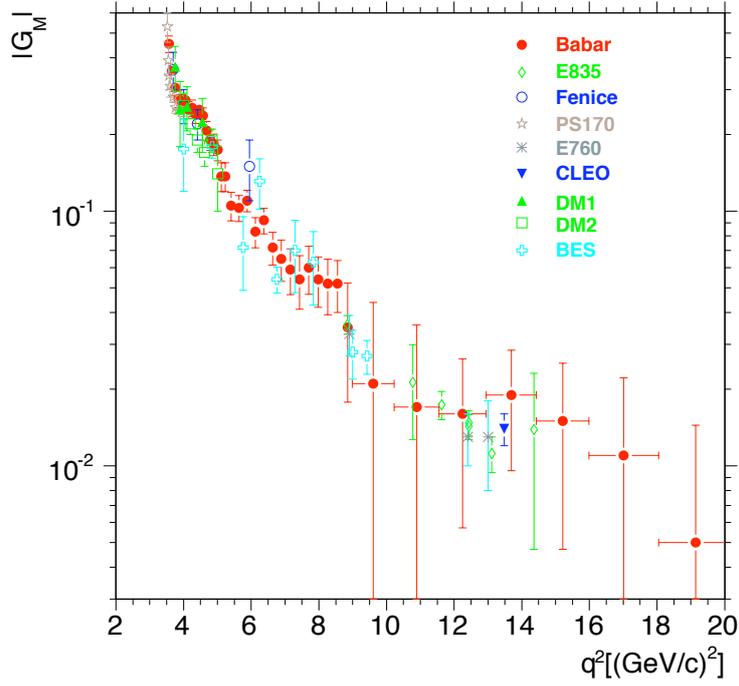

Figure 30: World data on the modulus $|G_M|$ of the time-like magnetic form factor extracted from different experiments from $\bar{p}p \to e^+e^-$, $e^+e^- \to \bar{p}p$ and $e^+e^- \to \bar{p}p\gamma$ experiments. The analysis assumes $|G_E| = |G_M|$.

The main experimental background for $\overline{P}ANDA$ in measuring $\bar{p}p \to e^+e^-$ is the reaction $\bar{p}p \to \pi^+\pi^-$, which is about 6 orders of magnitude bigger in cross section. A background rejection with a factor of $10^8$ using the particle-identification capabilities of $\overline{P}ANDA$ is required. To study if this is feasible and the sensitivity to the shape of the angular distribution after reconstructions an intensive Monte Carlo study was carried out. The angular distributions shown in Figure 31 were used as input into the Monte Carlo calculation in an event generator.

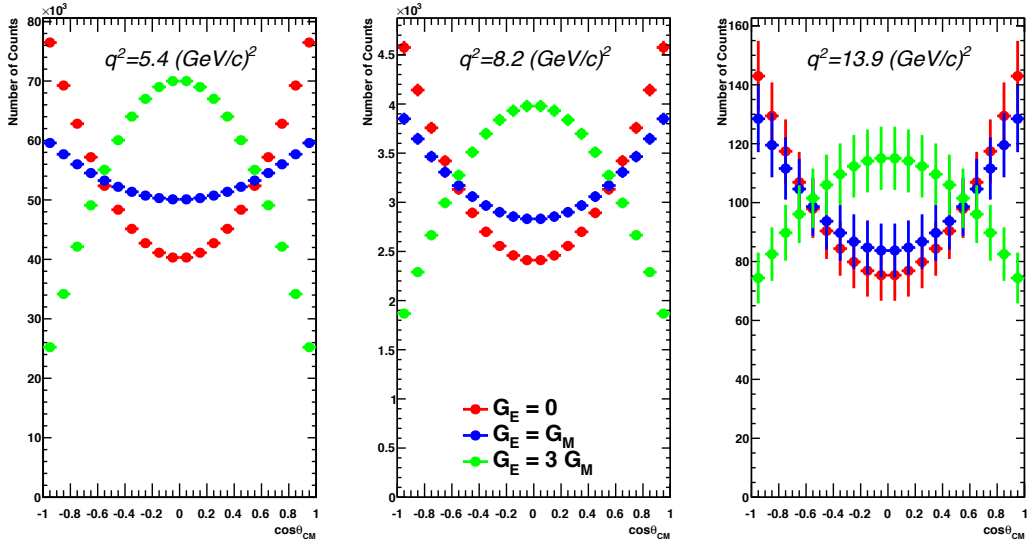

Figure 31: Event generator distributions (events before particle tracking and reconstruction) for $\bar{p}p \rightarrow e^+e^-$ for three different values of $q^2$. The three colors show the angular distribution in the center of mass frame (CMS) $|G_E| = 0$ (red), $|G_E| = |G_M|$ (blue) and $|G_E| = 3|G_M|$ (green). The error bars denote the statistical errors where no efficiency correction has been taken into account.

Four million signal events and $10^8$ pion background events have been analyzed in total. The extracted angular distribution after full reconstruction shows that the ratio $R=|G_E|/|G_M|$ can be determined since the angular distribution fits nicely to the generated Monte Carlo distribution (Figure 32), which had a specific $|G_E|/|G_M|$ ratio.

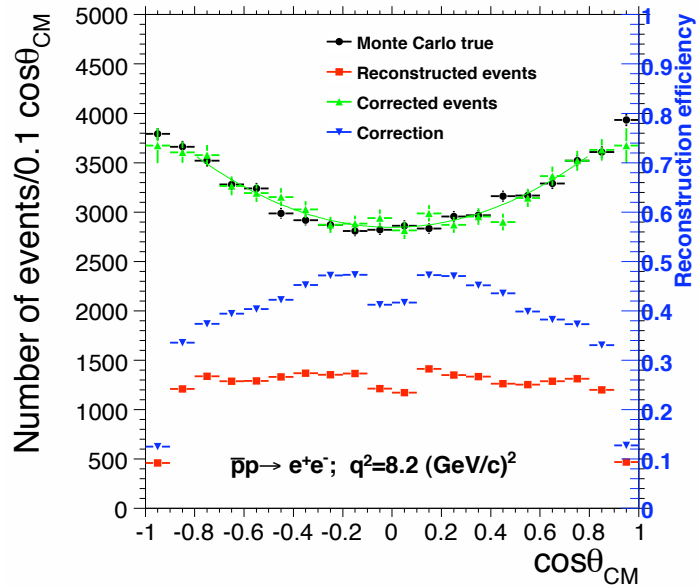

Figure 32: Angular distribution of the $e^+e^-$ pair in the CMS. Red symbols are the reconstructed $e^+e^-$ pairs and black symbols the Monte Carlo distribution of the event generator. The acceptance and reconstruction efficiency

correction for each point is given in blue. The corrected e$^+$e$^-$ pair distribution is shown in green and fits nicely the Monte Carlo distribution, which was generated for $R=|G_E|/|G_M|=1$.

As the result of this simulation, the expected error in $\overline{P}$ANDA for the ratio $R = |G_E|/|G_M|$ is shown in Figure 33. Data points from two previous experiments [87, 88] and three theoretical predictions for R [89] have also been included. It is obvious that $\overline{P}$ANDA will have no problems to distinguish between the models and will improve the overall data situation tremenduously.

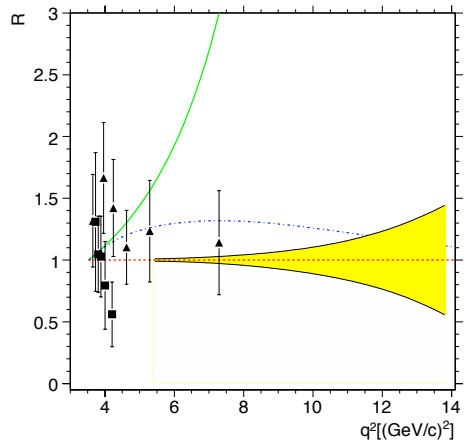

Figure 33: The expected error of $R=|G_E|/|G_M|$ as function of $q^2$ is shown as yellow band up to 14 (GeV/c)$^2$. Existing data points from previous experiments are shown as well as three different theory predictions (colored distributions).

In conclusion, $\overline{P}$ANDA can measure $R=|G_E|/|G_M|$ with unprecedented precision up to 14 (GeV/c)$^2$ without any assumptions regarding an exact luminosity determination. The values of $|G_E|$ and $|G_M|$ can be determined separately for this kinematical region. A precise luminosity measurement would allow $\overline{P}$ANDA to measure $R$ and the absolute and differential cross sections up to 22 (GeV/c)$^2$, which is the biggest range for a single experiment so far. Figure 34 shows the existing data set with the expected $\overline{P}$ANDA data points added.

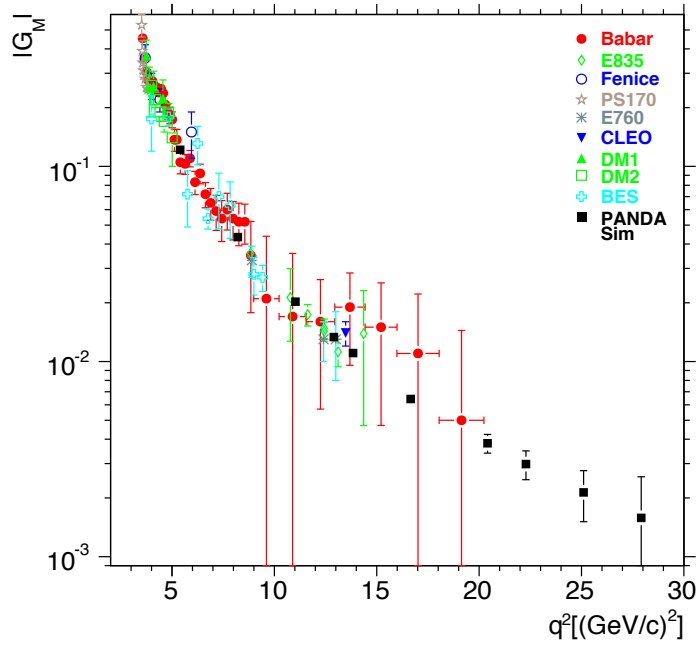

Figure 34: The existing world data on $|G_M|$ together with expected $\overline{P}$ANDA data. Each point corresponds to an integrated luminosity of 2 fb$^{-1}$.

**The Study of Transition Distribution Amplitudes in hard exclusive processes**

The scattering of virtual photons on a nucleon, with small momentum transfer squared $t$ between the baryons (i.e., in the forward region), can be described if factorization is possible by GPDs as already outlined (see Figure 26). The backward region is defined in terms of small momentum transfer squared $u$ between the baryon and the particle produced in the process. A factorized description of the backward deeply virtual exclusive reactions can be given in terms of Transition Distribution Amplitudes (TDAs). The kinematics require the exchange of 3 quarks in the $u$ channel. A comparison between the forward and backward processes is shown in Figure 35 [90].

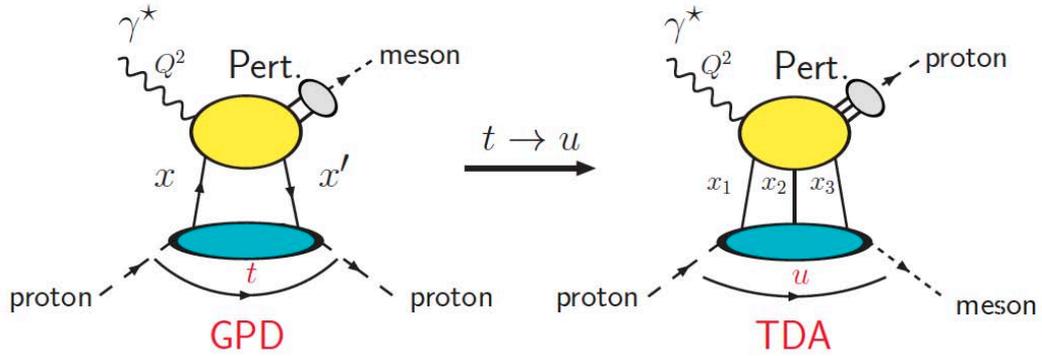

Figure 35: The two extreme but complementary limits of hard exclusive processes of deeply virtual Compton scattering (DVCS) can be described by GPDs in the forward region (left) and by TDAs in the backward region (right) [90].

In $\overline{\text{P}}\text{ANDA}$, the process $\bar{p}p \rightarrow \gamma^*\pi^0$ can be studied and involves the same TDAs as for backward electroproduction [90, 91] (see Figure 36).

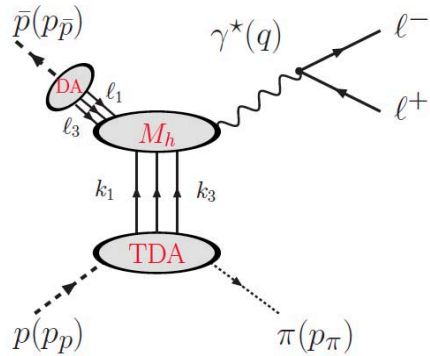

Figure 36: The production of a meson and a lepton-antilepton pair in the antiproton-proton annihilation process involves the same TDAs as for backward DVCS scattering [90].

The momentum transfer between the pion and the proton (or antiproton) should be small, such that the three quarks that are exchanged are collinear. Two different cases, shown in Figure 37, appear possible for studies [92]:

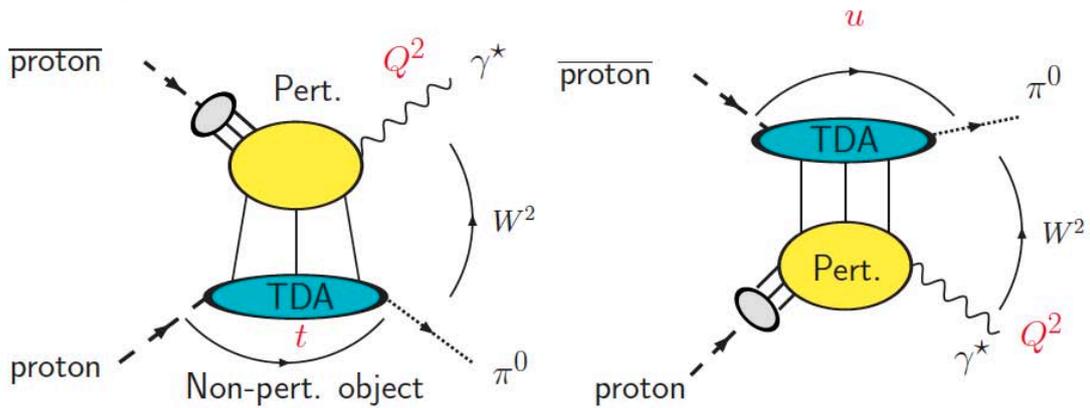

Figure 37: The production of a π⁰ and a virtual photon in the antiproton-proton annihilation process. The pion can be produced on the proton (left) or on the antiproton (right) [92].

From the predicted cross section of

$$\sigma^{l^+l^-\pi^0}(7 < Q^2 < 8 \text{ GeV}^2, W^2 = 10 \text{ GeV}^2, \Delta_T < 0.5 \text{ GeV}) \sim 100 \text{ fb} \quad [91]$$

$\overline{\text{P}}$ANDA can expect to detect roughly 100 events in 100 days of running for this kinematical region. The trigger would run in parallel to other physics programs.

The production of other light mesons like the η or the ρ could also be studied by $\overline{\text{P}}$ANDA. Even charmonium production is interesting in the context of TDAs, since the same TDAs should also appear in the process $\overline{\text{p}}\text{p} \rightarrow \text{J}/\psi + \pi^0$ (see Figure 38) [90].

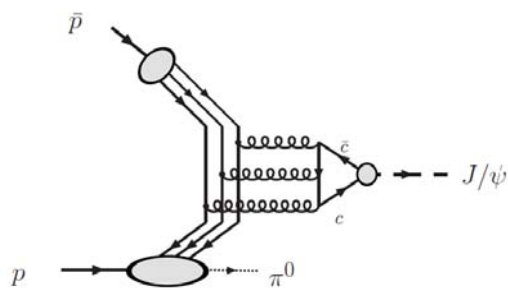

Figure 38: The production of a J/ψ together with a pion in the antiproton-proton annihilation process involves the same TDAs as the annihilation into a pion and two leptons [90].

To summarize: the three-dimensional structure of the proton can be tested in both the leptoproduction of particles in a lepton beam or in hadroproduction, e.g., in the antiproton-proton annihilation process. For leptoproduction, $Q^2$ is negative, while for the antiproton-proton annihilation process $Q^2$ is positive, which leads to different hard-scattering effects that can be studied systematically. The $\overline{P}$ANDA results on TDAs will yield the probability of finding a meson in the nucleon's three-dimensional picture.

## Hadronic Interaction Studies in $\overline{P}$ANDA

### Hypernuclear Physics in $\overline{P}$ANDA

Hypernuclei are nuclear systems in which one or more nucleons are replaced by one or more hyperons. A hyperon in a nucleus is not bound by the Pauli principle and unlike a neutron or proton, it can populate all possible nuclear states. It is not clear if, once implanted in a nucleus, the properties of a hyperon might change. Hyperons may also play a role in neutron stars. Certainly they provide a sensitive probe for the nuclear structure, which might be altered due to the presence of the embedded hyperon.

Nucleon-nucleon scattering was extensively studied in hadronic interactions. Similar studies using hyperons are impractical because of their short lifetimes. However, the spectroscopy of double hypernuclei will provide unique information on the hyperon-hyperon interaction.

Figure 39 shows our present knowledge on hypernuclei. Very few double hypernuclei have been seen so far.

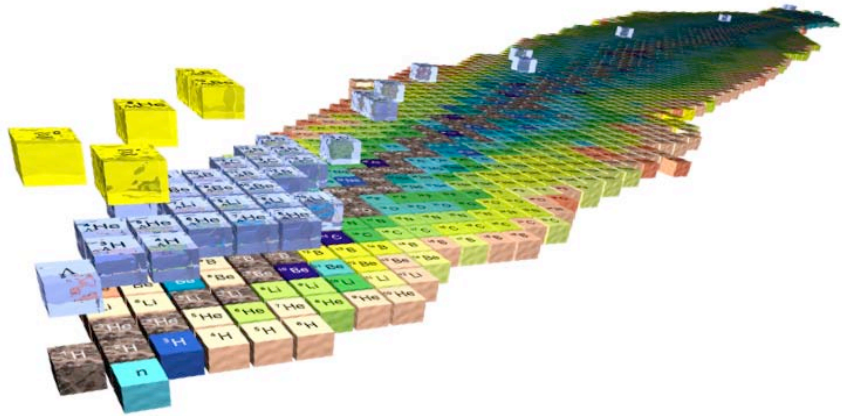

Figure 39: The present knowledge on hypernuclei. So far only single Λ-hypernuclei have been observed close to the valley of stability and only a few double hypernuclei events have been detected so far.

In the ground state, the hypernucleus can decay solely by the strangeness-changing weak interaction. Processes like ΛN → NN are allowed, while the decay with a meson Λ → Nπ is disfavored for heavier nuclei (see Figure 40) [93].

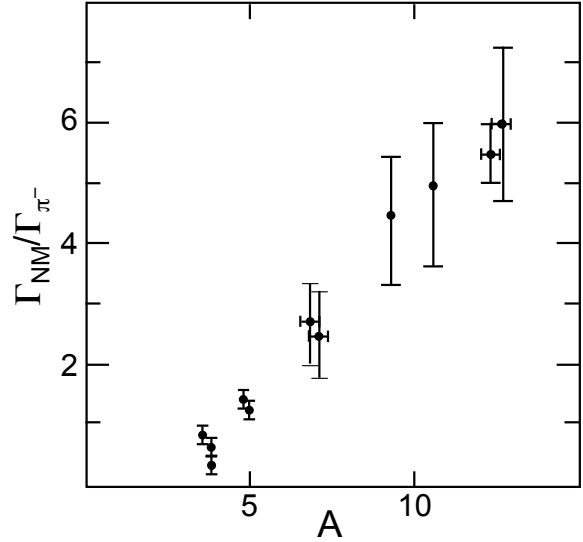

Figure: 40: The ratio of non-mesonic to mesonic hypernuclei decay widths as a function of the nucleus mass.

Double hypernuclei would even permit the study of $\Lambda\Lambda \to \Lambda N$ and $\Lambda\Lambda \to \Sigma N$ transitions [94, 95, 96]. The comparison between single and double hypernuclei should reveal valuable information on strangeness-nonconserving four-baryon interaction or non-mesonic weak decays.

For mesonic decays the one-meson exchange with strangeness and isospin does not play a role in the $\Xi$-N interaction, while only strange mesons must be considered in the $\Xi N$-$\Lambda\Lambda$ coupling. Only non-strange mesons are exchanged in the $\Lambda$-$\Lambda$ interaction [97, 98].

At the J-PARC accelerator, where hypernuclei are produced with kaon beams, only studies of the $\Xi$ hypernuclei seem feasible [99]. With $\overline{P}ANDA$, the physics of $\Omega^-$ atoms could come into reach. The $\Omega$ hyperon is the only baryon with a non-vanishing spectroscopic quadrupole moment, which is mainly determined by the one-gluon exchange contribution to the quark-quark interaction [100, 101].

Double hypernuclei could also be a doorway to producing the famous H-dibaryon [102, 103, 104]. This six-quark object (uuddss) has long been sought but not found. Inside a nucleus, the conditions might be different and H-dibaryons could appear.

While hypernuclei experiments traditionally use kaon or pion beams to produce the hypernuclei, the $\overline{P}ANDA$ experiment takes a completely different and unique experimental approach to hypernuclear physics using the production of baryon-antibaryon pairs in the antiproton-proton annihilation process. Negatively charged hyperons, e.g., from the process $\overline{p}p \to \Xi^-\overline{\Xi}^+$ close to threshold, have the chance to be slowed down and stopped in a secondary target and form $\Xi$ hypernuclei. The $\Xi$ hypernuclei will then be converted in a second step to ΛΛ double hypernuclei. This is illustrated in Figure 41. Kaons and $\bar{\Xi}^+$ particles serve as useful triggers for the reaction.

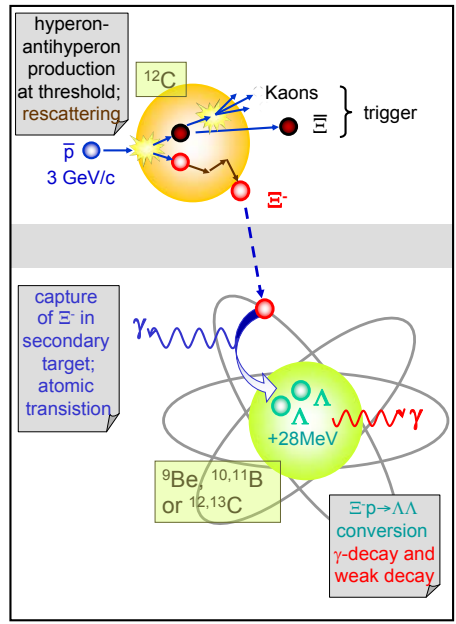

Figure 41: The double hyperon production process in $\overline{P}$ANDA

The $\overline{P}$ANDA detector will be built in a modular way to accommodate for the specific needs of hypernuclear physics. The backward end cap of the $\overline{P}$ANDA detector and the micro vertex detector (to avoid radiation damage) can be removed and a primary carbon target can be installed at the entrance of the central tracker of $\overline{P}$ANDA. A secondary target will be a sandwich arrangement of $^9$Be, $^{10}$B or $^{11}$B, or $^{12}$C or $^{13}$C absorbers to capture the $\Xi^-$, and of silicon detectors for the identification of the weak decay products. The isotropically emitted γ rays from the decay of the excited double hypernuclei will be detected by cluster arrays of germanium detectors at backward angles.

With a capture and conversion probability of 5-10% for the $\Xi^-$, a typical probability of 10% for a double-mesonic decay, and a data taking efficiency of 50%, we expect the spectroscopy of several tens of double hypernuclei per month while $\overline{P}$ANDA is running (see Figure 42). This has to be compared with less than 10 double hypernuclei events simply identified in the past 50 years.

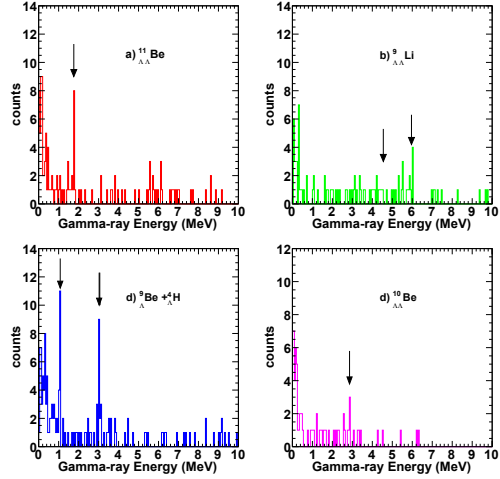

Figure 42: The expected γ transitions for single and double hypernuclei in a two weeks running period.

**In-medium modifications of charmed mesons**

It is believed that hadrons in nuclear matter could provide valuable information on the origin of hadron masses in the context of spontaneous chiral symmetry breaking in QCD by their modification due to chiral dynamics and partial restoration of chiral symmetry in a hadronic environment. In particular, attractive mass shifts rejecting the reduced quark pair condensate at finite density have been predicted for vector mesons [105, 106]. However, the connection between the chiral condensates and the nuclear density dependence of in-medium masses is not straightforward [107]. Experimentally the kinematics must be chosen so that the hadrons under investigation are either at rest or have a small momentum relative to the nuclear medium.

In addition to its mass, the second property of a hadron that could change in the nuclear medium is the width of the particle. Inside the medium, decay channels otherwise not allowed in the vacuum may open up and change the widths of hadrons.

The in-medium properties of charmonium or charmed hadrons could be studied experimentally for the first time with $\overline{\text{P}}$ANDA. In analogy to the observed $K/\overline{K}$ splitting in nuclear matter, theory predicts a splitting between the D and $\overline{\text{D}}$ mesons between 50 MeV [108, 109] and about 100 MeV at normal nuclear density [110, 111]. Since the high mass of the D mesons requires a high momentum in the antiproton beam to produce them, the conditions for observing in-medium effects seem unfavorable. The only way would be to find processes that would slow down the charmed hadrons inside the nuclear matter, but this requires more detailed theoretical studies. Certainly the search for in-medium properties of charmed mesons, despite the interesting physics involved, is not a day-one experiment for $\overline{P}ANDA$.

**D-meson interaction**

Although in-medium mass modifications are not easily accessible to $\overline{P}ANDA$, the possibility exists to study charmed meson *interaction* with nucleons inside nuclear matter. This could be achieved by tuning the antiproton energy to one of the higher-mass charmonium states that decays into open charm. The D mesons then have a chance to interact with the nucleons inside the target material. Since the D mesons are produced in pairs in the antiproton-nucleon annihilation process, the appearance of one of those D mesons can be used as tag to look for such reactions.

**The J/ψ dissociation cross section**

In hadron collisions, the J/ψ proceeds predominantly through gluon diagrams (e.g., gluon fusion) and provides a sensitive probe of the gluon structure function in the nucleon as well as its modification in nuclei. The J/ψ is also considered a leading indicator for the creation of hot-dense matter in heavy-ion collisions [112]. A significant suppression of a J/ψ signal in ultra-relativistic heavy-ion collisions would be considered a sign of quark-gluon plasma due to color screening. However, the interpretation of suppression requires the knowledge of the expected suppression due to the J/ψ dissociation in the hadronic environment. Knowledge of the momentum dependence of the dissociation would benefit the study of quark-gluon plasma.

In $\overline{P}ANDA$, the data for antiproton-nucleus collisions with momenta around 4 GeV/c can provide the J/ψ dissociation cross section with very little model dependence. At the proper momentum, the J/ψ production is a resonant formation process on a target nucleon. For different target nuclei (e.g., ranging from deuterium to gold), the J/ψ production cross section can be determined by varying the antiproton beam momentum over the J/ψ resonance profile. The width of this profile is essentially given by the nucleon momentum distribution in the target (Figure 43), which is sufficiently well known. The J/ψ is detected by its leptonic decays into electron or muon pairs. The attenuation of the J/ψ yield per effective target proton is then a direct measure of the J/ψN dissociation cross section, which can be deduced by a Glauber-type analysis.

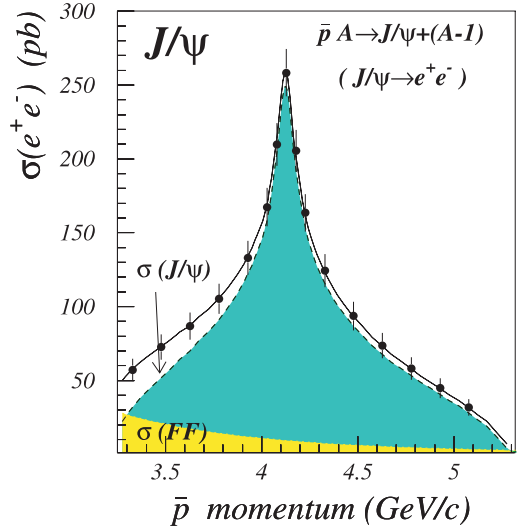

Figure 43: Monte Carlo simulation of the resonant J/ψ production cross section on nuclear proton with internal Fermi momentum distribution as function of the antiproton momentum [113].

Measurements of p̄N collisions can also be extended to antiproton momenta around 6.2 GeV/c, which allows for the ψ′ resonant production. Going to the ψ′ allows for the additional determination of the inelastic process $\psi'N \to J/\psi N$ [114]. This knowledge is also considered important for the interpretation of ultra-relativistic heavy-ion data. The yield for ψ′ production will be only a few percent compared to J/ψ production [115].

Picking up a J/ψ signal produced with a cross section of nb from an antiproton-nucleus collision with a total cross section of 1b seems like an impossible enterprise. The simulation of over 26 million UrQMD background events of the reaction p̄ $^{40}$Ca shows that in spite of these 9 orders of magnitude, no electron-positron pair has the same invariant mass as the J/ψ (Figure 44).

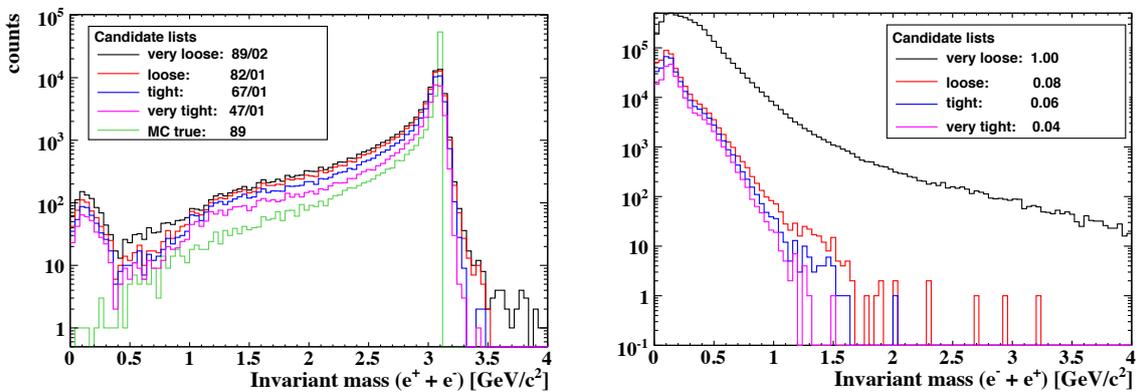

Figure 44: Reconstructed $e^+e^-$ invariant mass distributions for $\bar{p}\,^{40}$Ca annihilations. The signal is shown on the left, while the right figure shows background from UrQMD simulations.

A second source of background is the annihilation on a proton into two pions. The cross section is ~5 orders of magnitude higher for this process. Here, too, the excellent particle-identification capabilities of $\overline{\text{P}}$ANDA's electromagnetic calorimeter pay off. Figure 45 shows that a S/B ratio of greater than unity can be achieved and a determination of the J/$\psi$ dissociation cross section should be possible.

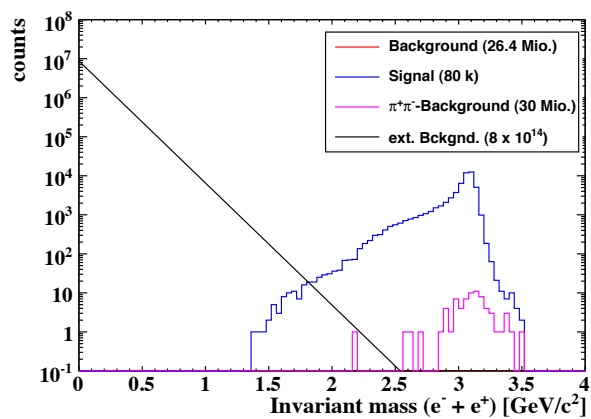

Figure 45: Invariant mass distribution of $e^+e^-$ pair candidates for signal (blue), UrQMD background (red), UrQMD background scaled to $10^{10}$ times the number of signal events (black) and $\pi^+\pi^-$ background (magenta) events.

**J/$\psi$-neutron cross sections**

The annihilations of antiprotons on a deuterium target could proceed as a two-step process. In the first step, the antiproton annihilates on the proton into a J/$\psi$, which then interacts with the spectator neutron. This should give direct access to the cross section of specific J/$\psi$n reactions [116].

**The antiproton-nuclear potential**

The antiproton-nuclear potential is unknown but predictions reach values of up to 700 MeV [117, 118, 119]. $\overline{\text{P}}$ANDA can study this, too, in a unique way. A high-energy antiproton will penetrate the nucleus and the potential will show up in the momentum of the forward and backward scattered antiprotons. An attractive antiproton potential will give the forward protons higher momenta than that of the incident ones. This should provide a very sensitive signature. The antiproton-nuclear potential reflects itself then in the missing-mass distribution of the fast-forward proton. Even the determination of the nuclear-antihyperon potential could be imagined in a similar way, after the incoming antiproton annihilates on a nuclear proton to produce an antihyperon-hyperon pair.

**Test of color transparency**

Another aspect of QCD might be investigated in the nuclear medium: color transparency. Color transparency manifests itself in the fact that hadrons produced in hard exclusive reactions have reduced interaction cross sections in a nucleus [120, 121, 122, 123]. This is a distinctive test for the underlying gauge theory, because a fundamental feature of gauge theory is that soft gluons decouple from the small color-dipole moment of the compact fast-moving color-singlet wave function configurations of the incident and final-state hadrons. $\bar{P}ANDA$ has the possibility to measure, not only on a proton target but also in a nuclear environment, specific reactions like

$$\bar{p}N \to \pi\pi \qquad \bar{p}N \to K\bar{K} \qquad \bar{p}N \to \bar{p}N$$

at various angles and energies that should be sensitive to color transparency. Varying the energy of the beam and changing the nuclear target should give the transparency ratio for various angle bins in the center-of-mass system with good precision.

## Flavor Physics and CP violation in $\bar{P}ANDA$

CP violation [124] has been established in neutral kaon and neutral B-meson decays. Studying D decays is the only chance to observe CP violation with an up-type quark. In the Standard Model, CP violation is suppressed for D mesons compared to kaons or B mesons. The advantage, of course, is that any signal in D decays would indicate new physics and the background from Standard Model processes is much reduced, making the observation cleaner.

The first signals for D mixing have been observed by several experiments; and even though no single experiment has a better value than 5σ, combining all results gives a 10σ evidence for mixing [125].

$\bar{P}ANDA$ has the advantage that it can reconstruct D mesons well. The open questions under investigation concern the production cross section and the background in the hadronic environment. Given the competition from LHCb and future B-factory projects, the production of $10^9$ D-meson pairs would be required. However, the flexibility of the $\bar{P}ANDA$ trigger, allows such searches for CP violation to run in parallel to other physics programs.

# The $\overline{\text{P}}$ANDA detector

**General considerations**

The $\overline{\text{P}}$ANDA detector has the huge benefit of being set up at a completely new facility, FAIR [126]. The detector can be optimized together with the accelerator for best performance. The design of both takes into account that it is costly in terms of time and money to produce antiprotons; $\overline{\text{P}}$ANDA is designed not as an experiment at an external beam but rather as an internal experiment in the HESR storage ring, so that the antiprotons not utilized for physics remain available. Previous storage-ring experiments suffered from a missing magnetic field, limiting the detection of charged particles. In $\overline{\text{P}}$ANDA, a solenoid surrounds the target in order to minimize holes in the acceptance of the detector. This is of particular importance in the forward region, where many of the particles go due the fixed-target arrangement. This is taken into account in the detector design by complementing the target spectrometer plus solenoid with a forward spectrometer plus dipole magnet. The dipole magnet, which is part of a chicane in the storage ring, bends the primary antiproton beam out of the way so that detectors and calorimeters can be placed in the 0° direction; thus the only hole in the acceptance is the beam pipe in backward direction, which is of less importance. The very nearly $4\pi$ coverage of the detector is essential for a partial-wave analysis that aims at the unambiguous determination of the quantum numbers of the resulting particles. An artist's view of the detector can be seen in Figure 46.

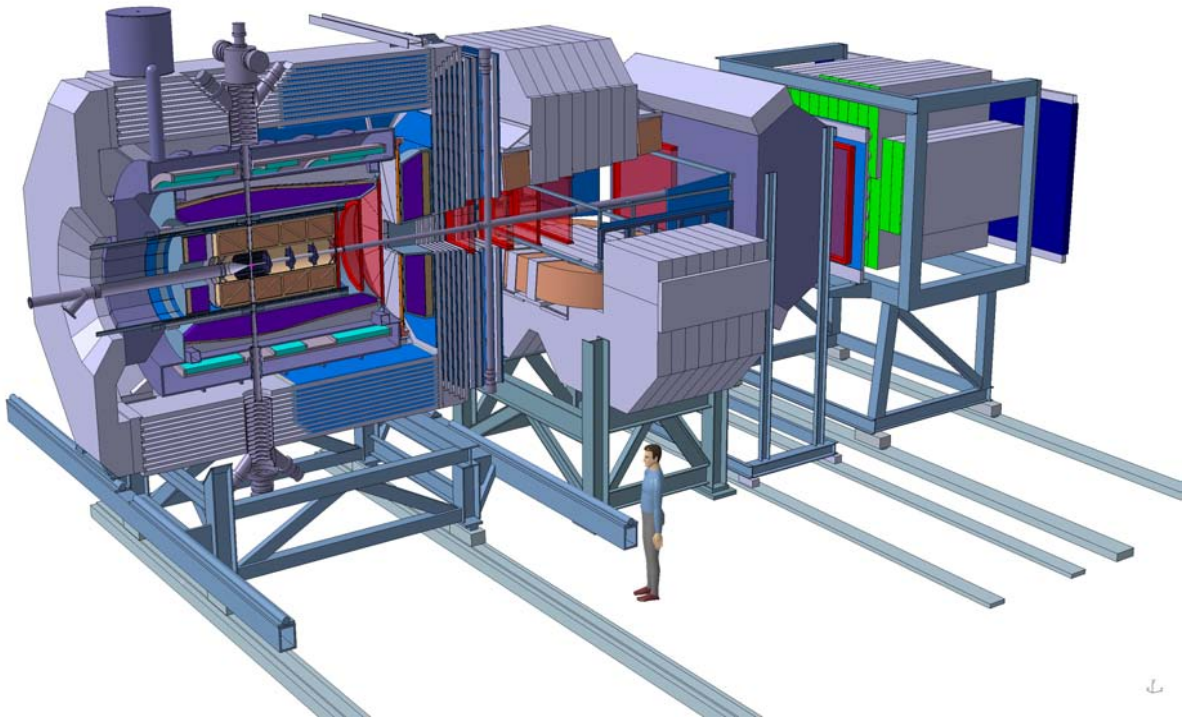

Figure 46: Artist's view of the $\overline{\text{P}}$ANDA experiment

The individual detector subsystems of $\overline{\text{P}}$ANDA allow for accurate charged-particle tracking, a high-resolution detection of photons, and sufficient particle identification at the highest interaction rates that could be taken by the detectors. Insights gained from the LHC experiments were taken into account and an advanced trigger system allows for an optimal utilization of the antiproton beam by allowing triggers for different physics programs to run in parallel.

The following sections first cover the HESR accelerator ring, which is an essential part for a successful $\overline{\text{P}}$ANDA experiment and will then go into a more detailed description of the individual detectors.

*The HESR Ring*

$\overline{\text{P}}$ANDA is located inside the HESR storage ring as an internal experiment. The advantage of a storage ring is that nearly all the antiprotons that are produced can be used for physics. This requires, of course, the constant cooling of the beam since the targets will constantly heat it.

The HESR ring is a racetrack-shaped ring with two long straight sections (see Figure 47).

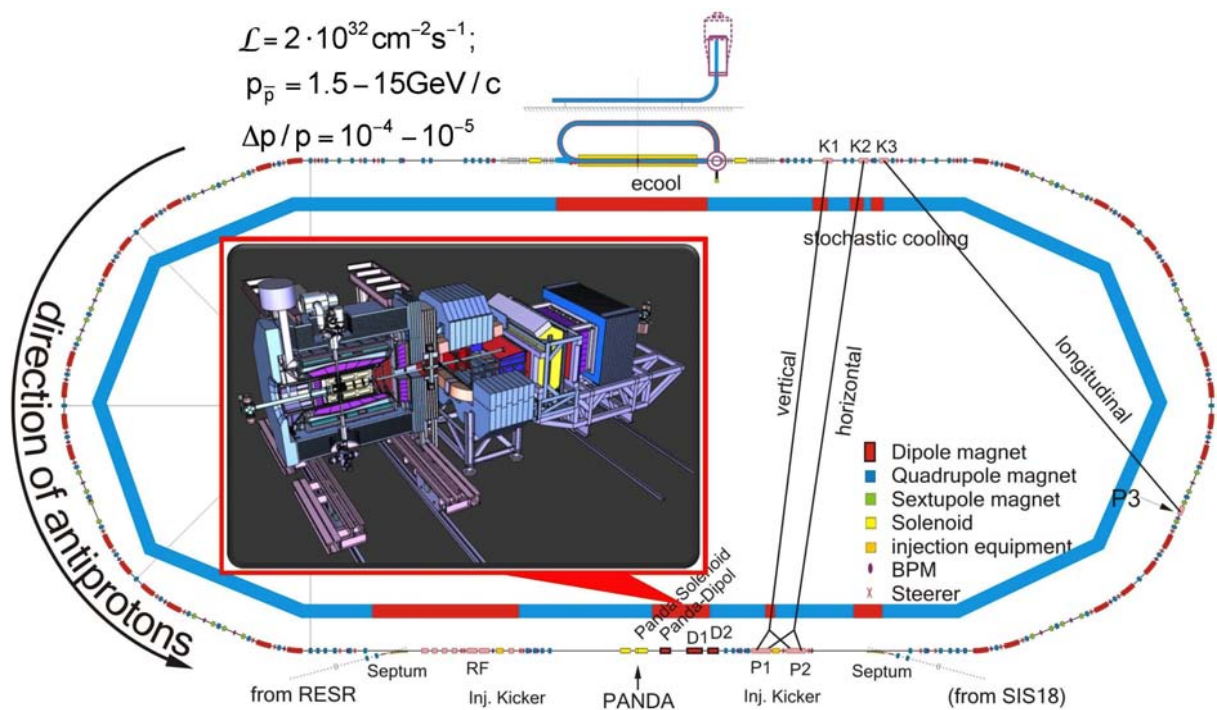

Figure 47: The HESR storage ring for antiproton physics at FAIR. The $\overline{\text{P}}$ANDA detector and the electron cooler are installed in the two straight sections.

The P̄ANDA experiment will be placed in one of the long straight sections. The other straight section will be used for an electron cooler. The electron cooling will be complemented by a stochastic cooling system. The antiprotons are injected with a momentum of 3.8 GeV/c into the HESR, which is capable of accelerating them up to 15 GeV/c or down to 1.5 GeV/c. Depending on the target or on the physics conditions, the beam lifetime is between 1500 and 7000 seconds. The ring can be used in a high-luminosity mode with beam intensities of up to $10^{11}$ antiprotons, or in a high-resolution mode with a momentum spread $\Delta p/p$ that can go down to a few times $10^{-5}$.

*The Target Spectrometer with the solenoid*

The target spectrometer surrounds the interaction point defined by the antiproton beam and the target. The detectors are arranged like the layers of an onion. A superconducting solenoid magnet provides a magnetic field of 2T for the charged-particle tracking. It is important to keep the whole detector as compact as possible, since the costs for the magnet or the calorimeter scale with its size. A side view of the target spectrometer is shown in Figure 48.

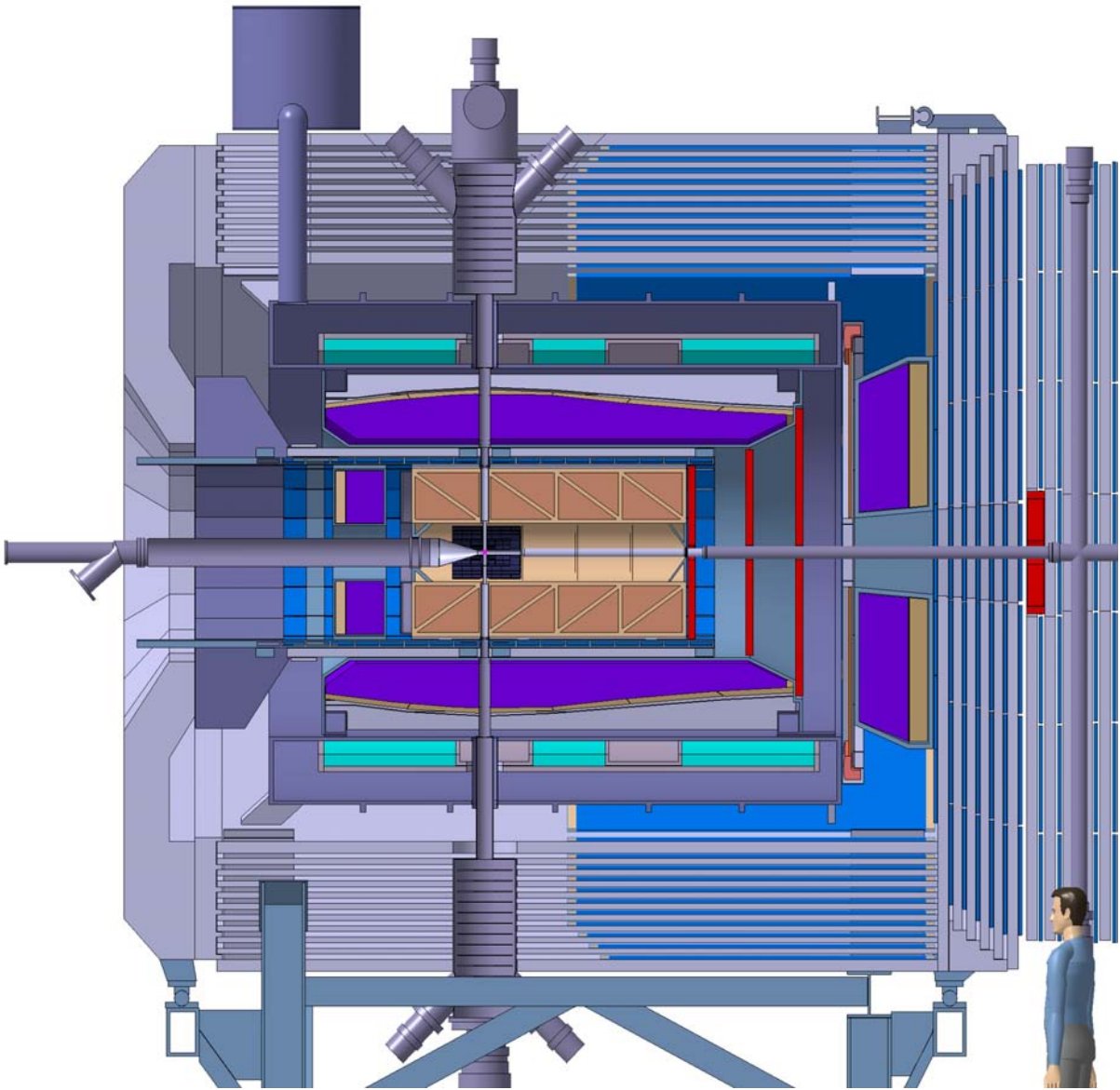

Figure 48: Side view of the target spectrometer

**The solenoid magnet**

In order to allow high-precision charged-particle tracking, the solenoid should provide a magnetic field of 2T with a homogeneity of better than 2% in the region of the charged-particle detectors. This can be achieved by a superconducting solenoid with an inner radius of 90 cm and a length of 2,80 m. The angular coverage with sufficient field for tracking is 5°/10°. If a time projection chamber serves as a central tracker in $\overline{P}ANDA$, the transverse component of the solenoidal field must be small for a uniform drift of the charges.

The coils of the magnet are placed outside the electromagnetic calorimeter in order to avoid the dead material in front of it. The iron yoke is segmented to incorporate muon chambers in a range telescope arrangement. The cryostat for the solenoid coils has two warm bores 100 mm in diameter, one above and one below the target position, to allow for insertion of internal targets.

**The targets**

The target inside a storage ring has to be tuned in terms of size and thickness. At present, two different, complementary techniques for the internal target are being developed: a cluster-jet target and a pellet target. Both techniques are capable of providing sufficient densities for hydrogen at the interaction point, but pose different challenges concerning their effect on the beam quality and the vertex definition of the interaction point. Beside hydrogen, internal targets of heavier gases—like deuterium, nitrogen or argon—can be made available in $\overline{\text{P}}$ANDA. In the case of non-gaseous nuclear targets, a thin wire target could be foreseen. The hypernuclear experiment that is planned requires a special target and a special arrangement for the backward half of the $\overline{\text{P}}$ANDA detector, which will be replaced in that case.

In a pellet target, a stream of frozen hydrogen micro-spheres, called pellets, traverses the antiproton beam perpendicularly. This technique was originally developed at the TSL laboratory in Uppsala for the WASA experiment [127] and is currently also in use at the WASA detector in Jülich. The pellets are between 15 and 40 μm and fall with a speed of about 60 m/s. A pellet tracking system utilizing a laser system and high-speed line-scan cameras is the subject of an R&D study, which should determine the pellet position and thus the interaction point to better than 100 μm. In case this does not prove to be feasible, another method will provide the exact location of the vertex origin. A single pellet becomes the vertex for more than a hundred nuclear interactions with antiprotons during the time a pellet traverses the beam. By averaging over many events, it will be possible to determine the position of individual pellets, using the resolution of the micro-vertex detector alone.

A disadvantage of a pellet target is the variation of the luminosity with the pellet flux. In this respect, clearly, a cluster-jet target with a homogeneous distribution of hydrogen atoms in the antiproton beam is superior. In a cluster-jet target, pressurized cold hydrogen gas expands into the vacuum through a Laval-type nozzle, which leads to a condensation of hydrogen molecules that form a narrow jet of hydrogen clusters. The density achieved so far is below that of a pellet target, however the expectations in current R&D are that the luminosity requirements for $\overline{\text{P}}$ANDA can be fulfilled. An additional advantage of a cluster-jet target is the homogeneous density profile and the possibility to focus the antiproton beam at highest phase-space density. The interaction point is defined transversely but has to be reconstructed longitudinally in the beam direction.

Both targets are connected to their source above the $\overline{\text{P}}$ANDA detector, and to a dump below the $\overline{\text{P}}$ANDA detector, by a thin pipe going through the whole detector and the magnet. A fixed cross with the beam pipe is installed at the interaction point.

**The micro vertex detector**

The micro vertex detector (MVD) of $\overline{\text{P}}$ANDA is the innermost tracking detector, mounted just outside the beam pipe. It is used to give high-resolution vertex points close to the interaction point and thus greatly improve the transverse momentum resolution. Also it allows the detection of secondary vertices from the decay of particles just outside the interaction point, which is e.g. the case for open-charm physics or hyperon decays.

The MVD consists of four layers of a barrel-type arrangement of silicon detectors, which is complemented by eight detector wheels in forward direction. The barrel has four layers with the innermost ones being pixel detectors of radiation-hard pixels with fast individual pixel readout. The two outer layers of the barrel are double-sided strip detectors. The two wheels in forward direction, which are closest to the.target consist also of pixel detectors Layers 2 through 6 in the forward direction combine pixels for the inner radius and strips for the outer one. The last two of the forward wheels consist entirely of strips and are placed further downstream to achieve better acceptance for hyperon decays, which have a longer decay length. Figure 49 shows the detector.

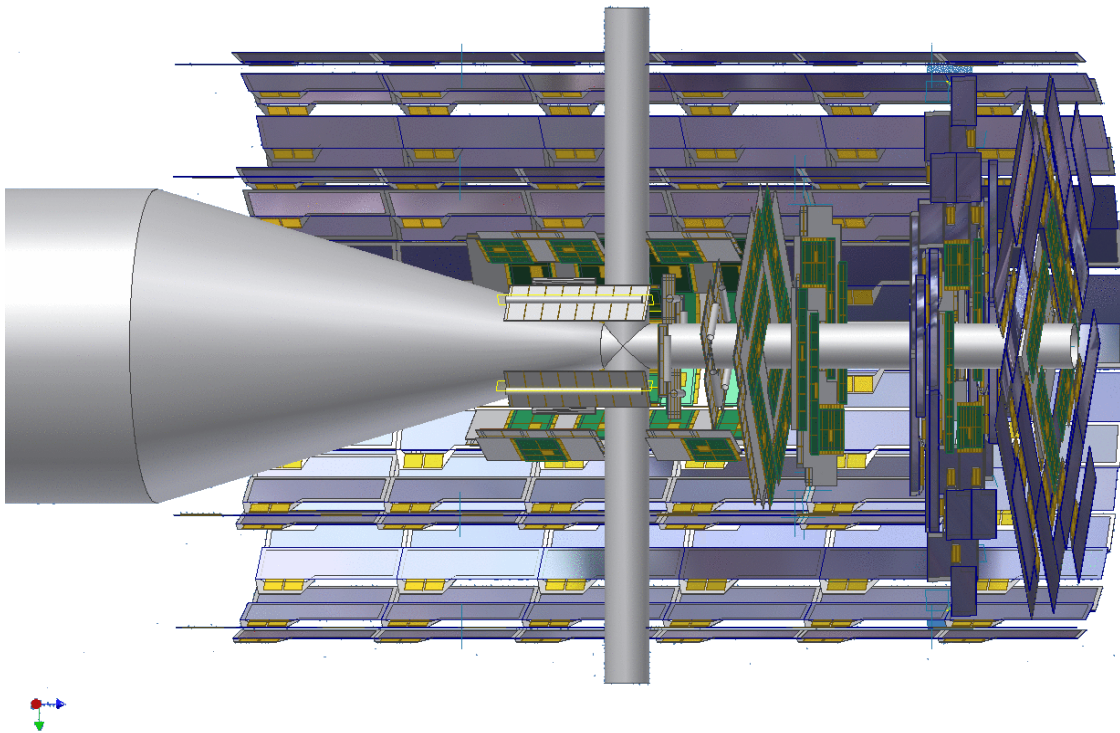

Figure 49: Artist's view of the $\overline{\text{P}}$ANDA micro vertex detector

The silicon waver thickness of 200 μm is currently based on the detectors in ATLAS and CMS at the LHC, but improving technologies might alter this in future.

**The central tracker**

Two options are currently under consideration for the main tracking device in $\overline{\text{P}}$ANDA: a straw tube tracker (STT) and a high-rate time projection chamber (TPC). Which of these options will proceed will be decided after the completion of the R&D program.

The STT being considered is built up from 30 μm thick aluminized mylar tubes with an anode wire inside; these are called straws. The straws, which are 10 mm in diameter, are operated at an overpressure of 1 bar, leading to a low-mass self-supporting structure. Inside $\overline{\text{P}}$ANDA the arrangement of straws will be kept in place by a thin and light frame. The low-mass construction adds only 1,3 % of a radiation length before the electromagnetic calorimeter.

The length of the detector is 150 cm and the inner and outer diameters are 15 cm and 42 cm. In total, 4200 straws are arranged in 24 planar layers mounted in a hexagonal shape around the micro vertex detector. The anode wire is made of 20 μm thick gold-plated tungsten. The foreseen gas mixture consists of Argon/$CO_2$ and the expected resolution in $x$ and $y$ coordinates is 150 μm. A $z$-resolution of 3 mm along the straws is achieved by tilting the 8 central layers.

A picture of the straw tracker is shown in Figure 50.

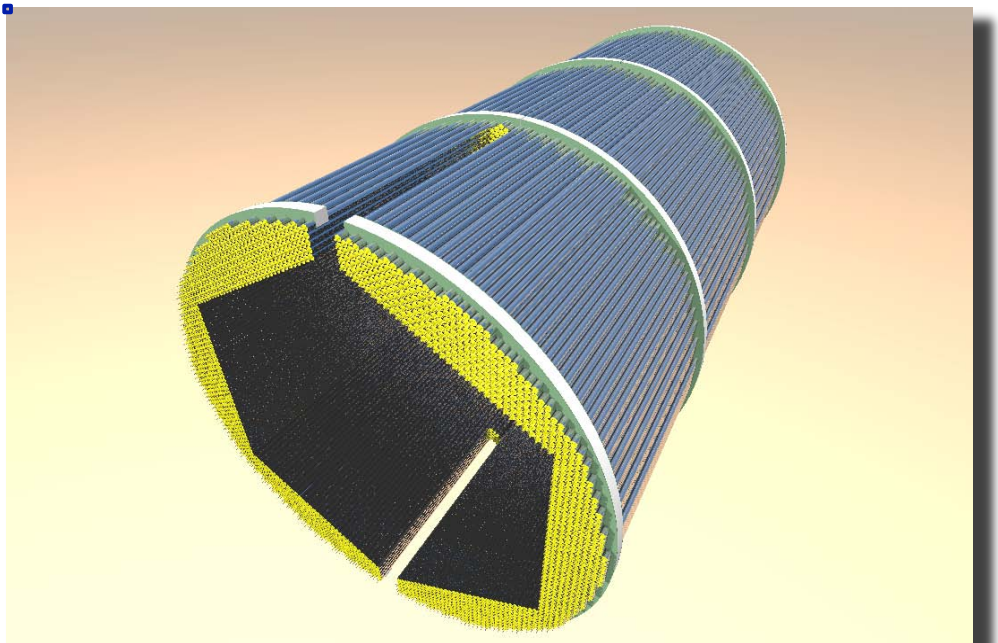

Figure 50: The $\overline{\text{P}}$ANDA central tracker with straw tubes

A TPC as the central tracker would have the advantage that it combines an accurate track resolution with an energy-loss measurement that provides additional particle identification, in particular for kaon-pion separation for momenta below 800 MeV/c. Since the TPC will have to work with large particle flows of up to 1000 events at the same time inside the TPC, the backflow of ions into the drift volume must be minimized to avoid space-charge build-up. This is done by so-called GEM foils, which allow the multiplication of the primary electrons by a factor of 1000 per foil. Due to the thinness of the foil of only 50 μm, the drift time of the ions is extremely short before they reach the cathode, so that space charge is no longer a problem. For the rare event that ions escape into the drift volume, the laser calibration system that is foreseen will measure the distortion of the field and allow for a later correction offline. The chamber is filled with $Ne/CO_2$ gas. The readout plane consists of 100,000 pads, each 2×2 $mm^2$ large. A disadvantage of the TPC is the very high data rate, which has to be handled by the data acquisition system (DAQ). A picture of the TPC can be seen in Figure 51.

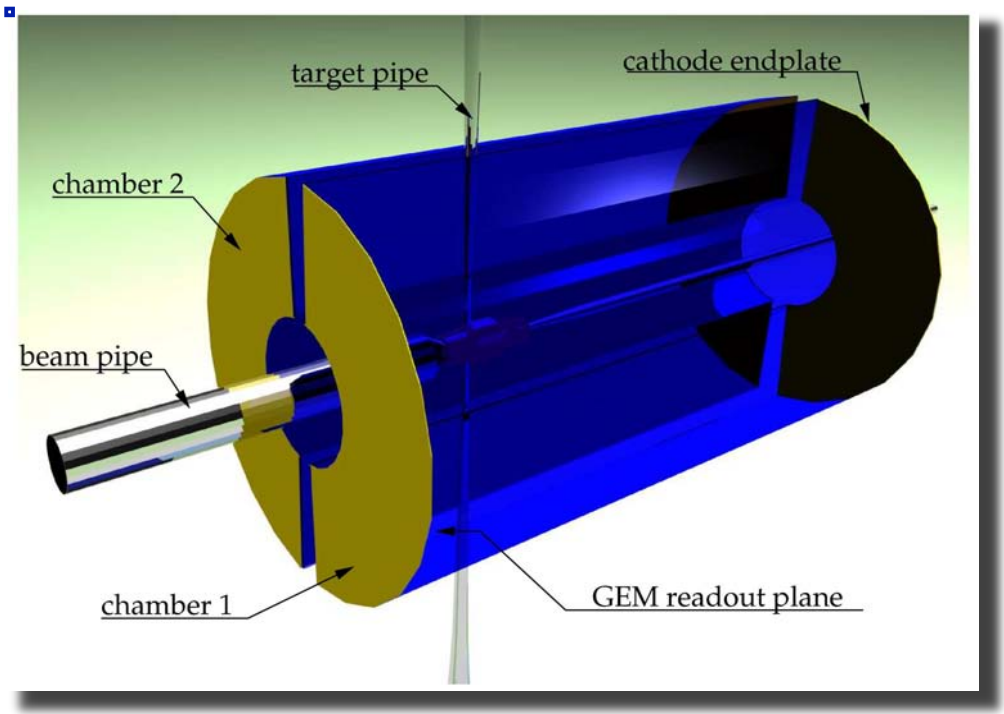

Figure 51: The central tracker realized as GEM TPC.

In the forward direction, three large planar chambers with GEM technology measure the tracks below 22°. Again, the GEM foils allow for a high-rate operation and for operation inside a magnetic field of 2T.

**Particle identification with Cherenkov detectors**

It is important for many physics channels to have active particle identification of charged particles, and this can be provided by Cherenkov detectors. An advanced technology, originally developed for the BaBar experiment at SLAC [128], uses the detection of internally reflected Cherenkov light in either quartz rods or in a disk quartz plane. The choice of the radiator material with a refractive index of 1.47 allows the discrimination of kaons and pions for momenta between 800 MeV/c up to 5 GeV/c. While the BaBar experiment used 11,000 photomultipliers to read out the light at the end of the rods, $\overline{\text{P}}$ANDA plans to use modern micro-channel plate photomultiplier tubes, which are insensitive to the magnetic field and also provide good time resolution for dispersion correction purposes. The disk, which is used in the forward direction in the target spectrometer, will be 2 cm thick and should help to identify charged particles between 5° and 22°.

**The electromagnetic crystal calorimeter**

In a hadronic environment, the detection of photons and leptons plays a major role in allowing for a nice, background-reduced trigger on interesting physics. The most prominent results were achieved, e.g, from antiproton annihilations into neutral channels by the Crystal Barrel experiment at LEAR. When it comes to the identification of charmonia, the lowest-lying charmonium states decay into electrons and a high-resolution electromagnetic calorimeter allows them to be reconstructed almost free from any background. The electromagnetic calorimeter is also the prime device for discriminating electrons from pions, which as mentioned earlier is an absolute necessity for certain physics processes.

The $\overline{\text{P}}$ANDA electromagnetic calorimeter is placed inside the superconducting coil and should be as compact as possible to keep the dimensions of the magnet as small as possible. The major challenges lie in the high counting rate—which reaches for the forward endplug up to 500 MHz—and the dynamic range of the photon energies—the $\overline{\text{P}}$ANDA apparatus must be able to detect photons with energies between 10 MeV and 15 GeV. It is obvious that the detector material must also be radiation-hard. A choice fulfilling most of the parameters is a scintillator made of $PbWO_4$ (PWO) crystals. These were developed originally for the CMS electromagnetic calorimeter. Since the requirements for $\overline{\text{P}}$ANDA are more stringent than for CMS in the detection of low-energy photons, it was necessary to improve the light output of the crystals. A successful research program implemented together with the manufacturer produced the so-called PWO II crystals, which provide a factor of 2 more light than the ones in CMS. A further factor of 4 in light ouput can be achieved by cooling the crystals to −25°, which reduces the thermal quenching of the luminescence process. PWO is one of the materials with the shortest radiation length, and $\overline{\text{P}}$ANDA covers 22 $X_0$ with its 20-cm-long crystals. The surface area pointing to a point slightly off the interaction point is 2×2 cm$^2$, which adapts to the Moliere radius for PWO. An energy resolution of $1\%/\sqrt{E/\text{GeV}}$, plus a constant term below 1%, has been achieved with such crystal matrices in test beams.

The use of photomultipliers is excluded because of the 2T magnetic field. $\overline{P}$ANDA has chosen to use Avalanche Photodiodes (APDs) for the photon detection. They combine good quantum efficiency, sufficient radiation hardness and provide enough external gain for the detection of 10 MeV photons A special type of rectangular 7×14 mm² APDs were developed together with the manufacturer and two APDs of this size cover the backward surface of the crystals nicely. Only for the forward end cap of the calorimeter, close to the beam pipe, do extremely high-counting rates require different photon detectors. The choice is to use vacuum photo triodes or tetrodes (VPTs or VPTTs). They work in principle like photomultipliers, but with only one or two dynodes they have only a weak dependence on the magnetic field.

The mechanics of the electromagnetic calorimeter poses a big challenge. It has to support 113,60 crystals arranged in a barrel configuration (see Figure 52), 3600 crystals in the forward end cap (see Figure 53), and 592 crystals in the backward end cap.

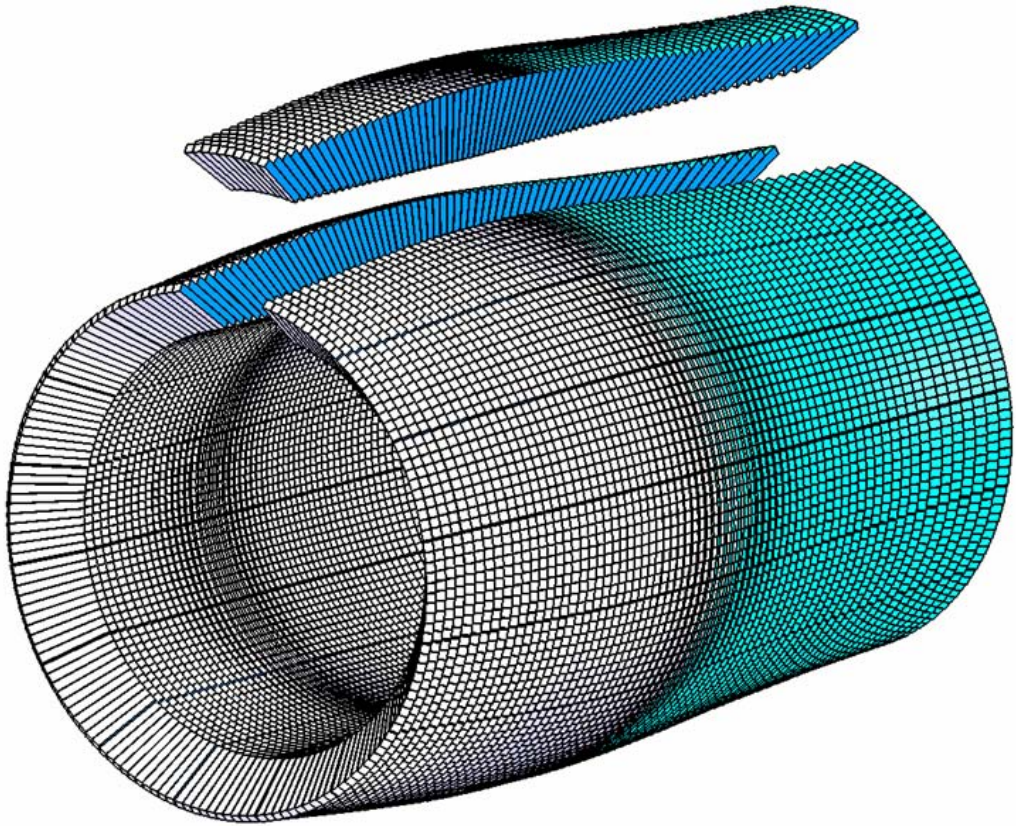

Figure 52: The barrel part of the $\overline{P}$ANDA electromagnetic calorimeter in the TS consists of ~11,360 PbWO$_4$ crystals arranged in 16 slices of 710 crystals each..

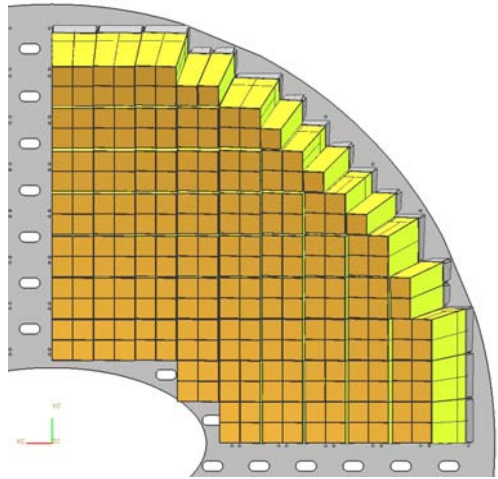

Figure 53: One quadrant of the forward endcap of the PANDA EMC in the TS

While the barrel calorimeter requires 11 different shapes of crystals for tight packing, only one type each is required for the two end caps, respectively. A major problem is the cooling of the crystals to −25°. In addition, a temperature stabilization to within ±0.1° is required to maintain the energy resolution of the crystals. A vacuum shield on the outside protects the crystals and the adjacent electronics from outside air, and active cooling is provided on both the rear and front sides of the crystals. Special attention has to be paid to the material budget of the shielding and cooling materials, and specially developed, extremely thin temperature sensors and humidity sensors constantly monitor the temperature and humidity inside the airtight volume.

**The muon detectors**

The detection of muons is not only important for the decay of charmonia into leptons, but also for studies of the Drell-Yan process or semileptonic open charm decays The main background for muon identification in a detector like $\overline{P}$ANDA comes from muons that arise from the decay of pions or kaons. Unlike the pions in the momentum range of $\overline{P}$ANDA , muons will easily penetrate the electromagnetic calorimeter and enter the iron yoke of the solenoid. This yoke is segmented and there are tracking detectors between the segments; this set-up not only detects kinks in the track from possible pion decays, but also acts as a range telescope. Drift tubes, as used for example by the COMPASS experiment at CERN, are foreseen as tracking detectors.

**The detector for hypernuclear physics**

Both, the production of hypernuclei and their spectroscopic investigation with high-resolution γ detectors require a change of the standard $\overline{P}$ANDA set-up. This is possible due to the modular design of the $\overline{P}$ANDA detector, which makes it possible to remove the backward end cap of the electromagnetic calorimeter and install a special target.

The production of double hypernuclei is done as a sequential two-step process. In the first step, a $\Xi\bar{\Xi}$ pair is produced in the antiproton annihilation on a proton in a nuclear target. Some of the $\Xi$ will be so slow that they can be absorbed in a secondary target close by. In the second step, a double hypernucleus is formed by a nuclear reaction in this secondary target. The secondary target is a 25−30 mm-thick sandwich structure of silicon micro strip detectors and absorbing material. The tracking by the silicon detectors can then detect the weak-decay cascade of the hypernucleus.

The γ spectroscopy will be done with existing germanium-array detectors, especially adapted to work in a magnetic field and an environment with high particle fluxes. The envisaged set-up inside the $\overline{\text{P}}$ANDA experiment is shown in Figure 54.

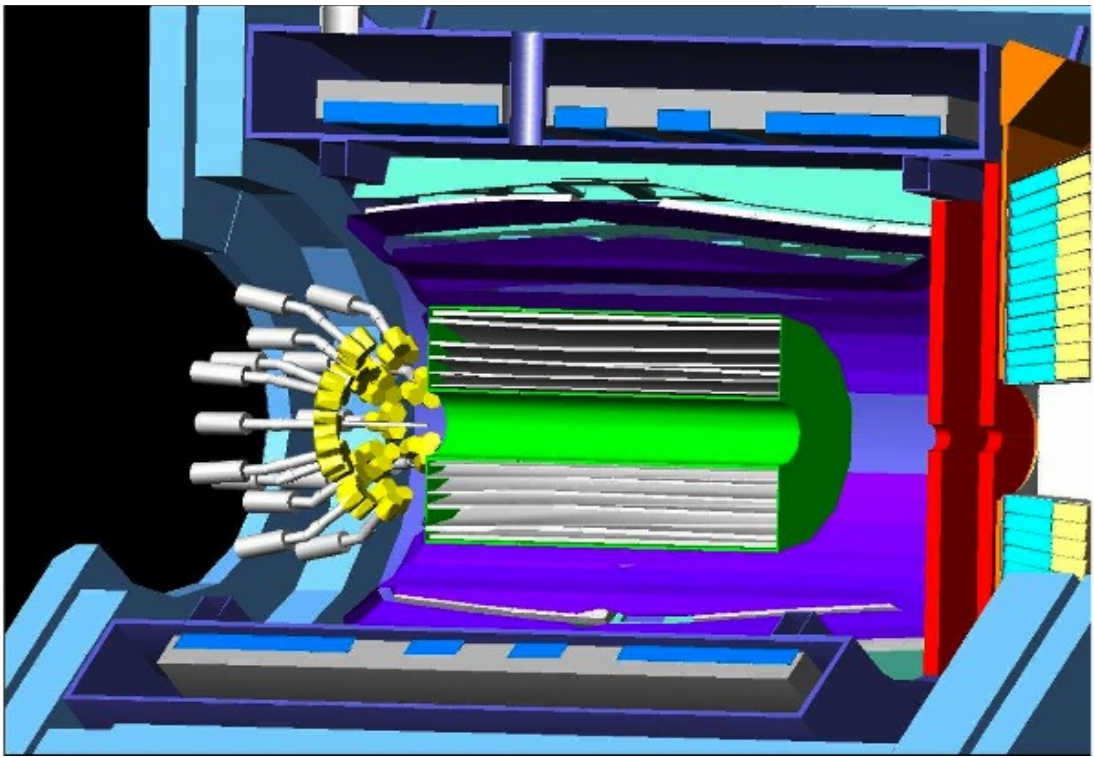

Figure 54: The modified $\overline{\text{P}}$ANDA detector for hypernuclear physics. The backward end cap and targets have been removed and are replaced by a retracted target and Ge detectors.

*The Forward Spectrometer with the dipole magnet*

The forward direction is important for $\overline{\text{P}}$ANDA because it is a fixed target experiment and most of the particles will go forward due to momentum conservation. The magnetic field for the charged-particle tracking in the forward direction is provided by a dipole magnet with a one-meter gap and an aperture of more than 2 meters. The magnet covers the entire angular acceptance of ±10° in the horizontal direction and only ±5° in vertical direction to avoid a huge gap between the two poles. The magnet will be a conventional magnet with a maximum bending power of 2 Tm. It will deflect the antiproton beam by 2.2° at its maximum momentum of 15 GeV/c, but this will be compensated by two correcting dipole magnets placed upstream and downstream of the $\overline{\text{P}}$ANDA experiment in the HESR ring.

**The forward trackers**

The momentum measurement for charged articles in the dipole field will be done by a set of six wire-chamber stations that are either small-cell-size drift chambers or straw tubes. Two of the stations will be in front of the dipole magnet, two inside and two behind.; the two stations inside the magnet will be able to track even particles with low momenta, whose tracks curl tightly in the magnetic field. Each station will contain three detection planes, one with vertical wires and the other two with wires inclined by +10° and −10°. Each chamber can therefore reconstruct a track individually, which might be important in the case of multi-track events. A central hole is foreseen in the chambers to accommodate the beam pipe. The expected momentum resolution of the tracking is, e.g., $\delta p/p = 0.2\,\%$ for 3 GeV/c protons.

**Forward particle identification**

A time-of-flight wall, made of plastic scintillator bars read out on both sides with fast photomultipliers, acts as a stop counter 7 m downstream from the target. The time resolution can be expected to be 50 ps. This will give a π/K separation up to momenta of 2.8 GeV/c and a K/p separation up to momenta of 4.7 GeV/c on a 3σ level. Higher momenta could be identified by a dual-radiator RICH (aerogel and gas). Space is foreseen for such an installation but no concrete plans have been made so far.

**The forward electromagnetic calorimeter and muon detectors**

The energy of forward-going photons and electrons will be measured in a Shashlyk calorimeter. The detection is based on a sandwich structure of lead and scintillators mounted in a block. The light is transported by wavelength-shifting fibers through the block to the outside-mounted photomultipliers. Previous experiments have achieved energy resolutions of $4\%/\sqrt{E}$. The electromagnetic calorimeter could be combined with a range tracking system for muons in the very forward direction. Such a system of interchanging detectors and absorbers is very similar to the muon trackers used in the target spectrometer.

**The luminosity monitor**

In order to determine the absolute cross section for certain physics processes being important for $\overline{\text{P}}$ANDA, the time-integrated luminosity has to be measured. One way to determine the absolute cross section is to measure the count rate of a reaction where the cross section is well-known. In the case of $\overline{\text{P}}$ANDA, we plan to take elastic antiproton-proton scattering as the reference channel and measure the differential cross section, which is dominated by Coulomb scattering for very low angles. The electromagnetic amplitude can be calculated precisely and hence the total cross section and the luminosity can be determined without measuring the inelastic rate.

The tracks of the scattered antiprotons must be determined as precisely as possible for scattering angles between 3 and 8 mrad to achieve the goal. This will be done by a set of four planes of double-sided silicon strip detectors situated 10–12 meters downstream of the $\bar{P}$ANDA target. Roman pots may be implemented to retract the detectors during beam development, to avoid radiation damages.

*The Data Acquisition System of $\bar{P}$ANDA*

The $\bar{P}$ANDA data acquisition system (DAQ) is extremely challenging. The data rate of $2 \times 10^7$ events per second is high and the detectors have very different response times, with fast scintillators and a very slow TPC at the two extremes. Given the diversity of the physics program, with very different triggers in parallel, the traditional approach of a hierarchical hardware trigger has to be abandoned.

In the $\bar{P}$ANDA experiment each of the detectors is a self-triggering entity. Intelligent front-end modules within each of these sub-systems deal with the data, extracting and transmitting the relevant information for a hit to the next level in the DAQ. $\bar{P}$ANDA plans to utilize field programmable gate arrays (FPGAs) and high-speed serial links for this purpose. Using programs instead of hardwiring the trigger delivers the flexibility needed for each particular physics purpose. Compute nodes at higher levels of the DAQ system then enable more sophisticated triggers. To ensure that the events are properly combined into an event, a precise time stamp marks each of the detector hits. The recorded data rate will be on the order of 100–200 MB/s, leading to a total data volume for $\bar{P}$ANDA of approximately 1 PB/year. It will be a huge challenge to analyze these data, but without a doubt the results will deepen our understanding of the strong interaction tremendously.

## Acknowledgements

The author acknowledges support from the German Ministry for Research and Education (BMBF) and the GSI and Jülich Helmholtz centers. He would like to thank his colleagues from the $\bar{P}$ANDA collaboration for lots of physics ideas and discussions and Michelle Mizuno for comments on the manuscript.